\documentclass[12pt]{article}

\usepackage{amsmath}
\usepackage{amssymb}
\usepackage{amsthm}
\usepackage{latexsym}
\usepackage{color}
\usepackage{graphicx}
\usepackage{color}

\DeclareSymbolFont{calletters}{OMS}{cmsy}{m}{n}
\DeclareSymbolFontAlphabet{\mathcal}{calletters}

\def\be{\begin{eqnarray}}
\def\ee{\end{eqnarray}}

\def\b*{\begin{eqnarray*}}
\def\e*{\end{eqnarray*}}

\newtheorem{Theorem}{Theorem}[section]
\newtheorem{Definition}[Theorem]{Definition}
\newtheorem{Proposition}[Theorem]{Proposition}

\newtheorem{Assumption}[Theorem]{Assumption}

\newtheorem{Lemma}[Theorem]{Lemma}

\newtheorem{Remark}[Theorem]{Remark}
\newtheorem{Example}[Theorem]{Example}

\makeatletter \@addtoreset{equation}{section}

\usepackage{color}

\def \E{\mathbb{E}}

\def \L{\mathbb{L}}
\def \P{\mathbb{P}}

\def \R{\mathbb{R}}

\def \M{\mathbf{M}}
\def \N{\mathbb{N}}

\def\Cc{{\cal C}}
\def\Dc{{\cal D}}

\def\Mc{{\cal M}}
\def\Nc{{\cal N}}

\def\Pc{{\cal P}}

\def \eps{\varepsilon}

\def \0{\mathbf{0}}

\def\1{{\bf 1}}
\def \proof{{\noindent \bf Proof.\quad}}

\def \no{\noindent}
\def \ep{\hbox{ }\hfill{ ${\cal t}$~\hspace{-5.1mm}~${\cal u}$}}

\title{An Explicit Martingale Version of Brenier's Theorem\thanks{The authors are grateful to Mathias
Beiglb\"{o}ck and Xiaolu Tan for fruitful comments.}}

\author{Pierre Henry-Labord\`ere \thanks{Soci\'et\'e G\'en\'erale, Global Market Quantitative Research,
pierre.henry-labordere@sgcib.com}
        \and Nizar Touzi\thanks{Ecole Polytechnique Paris, Centre de Math\'ematiques Appliqu\'ees,
        nizar.touzi@polytechnique.edu} }

\date{\today}

\begin{document}

\maketitle

\begin{abstract}
By investigating model-independent bounds for
exotic options in financial mathematics, a martingale version of the Monge-Kantorovich mass transport problem was
introduced in \cite{BeiglbockHenry-LaborderePenkner,GalichonHenry-LabordereTouzi}. In this paper, we extend the
one-dimensional Brenier's theorem to the present martingale version. We provide the explicit martingale
optimal transference plans for a remarkable class of coupling functions corresponding to the lower
and upper bounds. These explicit extremal probability measures coincide with the unique left and
right monotone martingale transference plans, which were introduced in
\cite{BeiglbockJuillet} by suitable adaptation of the notion of cyclic monotonicity. Instead, our
approach relies heavily on the (weak) duality result stated in \cite{BeiglbockHenry-LaborderePenkner},
and provides, as a by-product, an explicit expression for the corresponding optimal
semi-static hedging strategies. We finally provide an extension to the multiple marginals case.
\end{abstract}

\section{Introduction}

Since the seminal paper of Hobson \cite{Hobson}, an important
literature has developed on the topic of robust or model-free
superhedging of some path dependent derivative security with payoff
$\xi$, given the observation of the stochastic process of some
underlying financial asset, together with a class of derivatives.
See \cite{BrownHobsonRogers, Cousot1, Cousot2, CoxHobsonObloj,
CoxObloj1, CoxObloj2, CoxWang, DavisHobson, DavisOblojRaval,
HobsonKlimmek, HobsonPedersen, OberhauserDosreis} and the survey
papers of Obl\'{o}j \cite{Obloj} and Hobson \cite{Hobson-LNM}. In
continuous-time models, these papers mainly focus on derivatives
whose payoff $\xi$ is stable under time change.
 Then, the key-observation was that, in the idealized context where all $T-$maturity European calls and puts, with
 all possible strikes, are available for trading, model-free superhedging cost of $\xi$ is closely related to
 the Skorohod Embedding problem. Indeed, the market prices of all $T-$maturity European calls and puts with all possible strikes allow to recover the marginal distribution of the underlying asset price at time $T$.

Recently, this problem has been addressed via a new connection to
the theory of optimal transportation, see
\cite{BeiglbockHenry-LaborderePenkner, GalichonHenry-LabordereTouzi,
Henry-LabordereOblojSpoidaTouzi,
AcciaioBeiglbockPenknerSchachermayerTemme,AcciaioBeiglbockPenknerSchachermayer,
DolynskiSoner1, DolynskiSoner2}. Our interest in this paper is on
the formulation of a Brenier Theorem in the present martingale
context. We recall that the Brenier Theorem in the standard optimal
transportation theory states that the optimal coupling measure is
the gradient of some convex function which identifies in the
one-dimensional case to the so-called Fr\'echet-Hoeffding coupling
\cite{Brenier}. A remarkable feature is that this coupling
is optimal for the class of coupling cost functions satisfying the
so-called Spence-Mirrlees condition.

We first consider the one-period model. Denote by $X,Y$ the prices
of some underlying asset at the future maturities $0$ and $1$,
respectively. Then, the possibility of dynamic trading implies that
the no-arbitrage condition is equivalent to the non-emptyness of the
set $\Mc_2$ of all joint measures $\P$ on $\R_+\times\R_+$ satisfying
the martingale condition $\E^\P[Y|X]=X$. The model-free subhedging and superhedging costs of some derivative security with payoff
$c(X,Y)$, given the marginal distributions $X\sim\mu$ and
$Y\sim\nu$, is essentially reduced to the martingale transportation
problems:
 \b*
 \inf_{\P\in\M_2(\mu,\nu)}\E^\P[c(X,Y)]
 &\mbox{and}&
 \sup_{\P\in\M_2(\mu,\nu)}\E^\P[c(X,Y)],
 \e*
where $\M_2(\mu,\nu)$ is the collection of all probability measures
$\P\in\Mc_2$ such that $X\sim_\P\mu$, $Y\sim_\P\nu$. Our main
objective is to characterize the optimal coupling measures which
solve the above problems. This provides some remarkable extremal
points of the convex (and weakly compact) set $\M_2(\mu, \nu)$. In
the absence of marginal restrictions, Jacod and Yor \cite{JacodYor}
(see also Jacod and Shiryaev \cite{JacodShiryaev}, Dubins and
Schwarz \cite{dub}, for the discrete-time setting) proved that a
martingale measure $\P \in {\cal M}_2$ is extremal if and only if
$\P$-local martingales admit a predictable representation. In the
present one-period model, such extremal points of ${\cal M}_2$
consist of binomial models. For a specific class of coupling
functions $c$, the extremal points of the corresponding martingale
transportation problem turn out to be of the same nature, and our
main contribution in this paper is to provide an explicit
characterization.

Our starting point is a paper by Hobson and Neuberger
\cite{HobsonNeuberger} who considered the specific case of the
coupling function $c(x,y):=|x-y|$, and provided a complete explicit
solution of the optimal coupling measure and the corresponding
optimal semi-static strategy. In a recent paper, Beiglb\"{o}ck and
Juillet \cite{BeiglbockJuillet} address the problem from the
viewpoint of optimal transportation. By a convenient extension of
the notion of cyclic comonotonicity, \cite{BeiglbockJuillet}
introduce the notion of left-monotone transference plan. They also
introduce the notion of left-curtain as a left-monotone transference
plan concentrated on the graph of a binomial map. The remarkable
result of \cite{BeiglbockJuillet} is the existence and uniqueness of
the left-monotone transference plan which is indeed a left-curtain,
together with the optimality of this joint probability measure for
some specific class $\cal C_{\rm BJ}$ of coupling payoffs $c(x,y)$.
Notice that the coupling measure of \cite{HobsonNeuberger} is not a
left-curtain, and $\cal C_{\rm BJ}$ does not contain the coupling
payoff $|x-y|$.

As a main first contribution, we provide an explicit description of the left-curtain $\P_*$ of \cite{BeiglbockJuillet}. Then, by using the weak duality inequality,

- we provide a larger class $\Cc\supset\Cc_{\rm BJ}$ of payoff functions for which $\P_*$ is optimal,

- we identify explicitly the solution of the dual problem which consists of the optimal semi-static superhedging strategy,

- as a by-product, the strong duality holds true.

Our class $\Cc$ is the collection of all smooth functions
$c:\R\times\R\longrightarrow\R$, with linear growth, such that
$c_{xyy}>0$. We argue that this is essentially the natural class for
our martingale version of the Brenier Theorem.

We next explore the multiple marginals extension of our result. In the context of the finite discrete-time model, we provide a direct extension of our result which applies to the context of the discrete monitored variance swap. This answers the open question of optimal model-free upper and lower bounds for
this derivative security.

The paper is organized as follows. Section \ref{sect:Brenier} provides a quick review of the Brenier Theorem in the standard one-dimensional optimal transportation problem. The martingale version of the Brenier Theorem is reported in Section \ref{sect:mainresults}. We next report our extensions to the multiple marginals case in Section \ref{sect:multimarginals}. Finally, Section \ref{sect:proofs} contains the proofs of our main results.

\section{The Brenier Theorem in One-dimensional Optimal Transportation}
\label{sect:Brenier}

\subsection{The two-marginals optimal transportation problem}
\label{sect-OT}

Let $X$, $Y$ be two scalar random variables denoting the prices
of two financial assets at some future maturity $T$. The pair
$(X,Y)$ takes values in $\R^2$, and its distribution is defined by
some probability measure $\P\in{\cal P}_{\R^2}$, the set of all
probability measures on $\R^2$. For the purpose of the present
financial application, the measures have support on $\R_+^2$. For
the sake of generality, we consider however the general case.

We assume that $T-$maturity European call options, on each asset
and with all possible strikes, are available for trading  at
exogenously given market prices. Then, it follows from Breeden and
Litzenberger \cite{bre} that the marginal distributions of $X$ and
$Y$ are completely determined by the second derivative of the
corresponding (convex) call price function with respect to the
strike. We shall denote by $\mu$ and $\nu$ the implied marginal
distributions of $X$ and $Y$, respectively, $\ell^\mu,r^\mu$,
$\ell^\nu,r^\nu$ the left and right endpoints of their supports, and
$F_\mu$, $F_\nu$ the corresponding cumulative distribution
functions.

By definition of the problem the probability measures $\mu$ and
$\nu$ have finite first moment:
 \begin{eqnarray}\label{firstmoment}
 \int |x|\mu(dx)+\int|y|\nu(dy)
 &<&
 \infty,
 \end{eqnarray}
and although the supports of $\mu$ and $\nu$ could be restricted to
the non-negative real line for the financial application, we shall
consider the more general case where $\mu$ and $\nu$ lie in ${\cal
P}_\R$, the collection of all probability measures on $\R$.

We consider a derivative security defined by the payoff $c(X,Y)$ at
maturity $T$, for some upper semicontinuous function
$c:\R^2\longrightarrow\R$ with linear growth. This condition could
be replaced by
 \be\label{othercond}
 c(x,y) \leq \varphi(x)+\psi(y)
 &\mbox{for some}&
 \varphi,\psi:\R\longrightarrow\R,~~
 \varphi^+ \in \L^1(\mu), \psi^+\in \L^1(\nu).
 \ee
The model-independent upper bound for this payoff, consistent with
vanilla option prices of maturity $T$, can then be framed as a
Monge-Kantorovich (in short MK) optimal transport problem:
 \b*
 P^0_2(\mu,\nu)
 :=
 \sup_{\P \in{\cal P}_2(\mu,\nu)}
 \E^{\P}\big[c(X,Y)\big]
 &\mbox{where}&
 {\cal P}_2(\mu,\nu)
 :=
 \big\{\P\in{\cal P}_{\R^2}:X\sim_\P\mu
                              ~\mbox{and}~
                              Y\sim_\P\nu
 \big\},
 \e*
where, for the sake of simplicity, we have assumed a zero interest
rate. This can easily be  relaxed by considering the forwards of $X$
and $Y$. Notice that $c(X,Y)$ is measurable by the upper
semicontinuity condition on $c$, and is integrable by the linear
growth condition on $c$ together with (\ref{firstmoment}) (or the
condition (\ref{othercond})).

In the original optimal transportation problem as formulated by
Monge, the above maximization problem was restricted to the
following subclass of measures.

\begin{Definition}
A probability measure $\P\in\Pc_2(\mu,\nu)$ is called a transference
map if $\P(dx,dy):=\mu(dx)\delta_{\{T(x)\}}(dy)$, for some
measurable map $T:\R\longrightarrow\R$.
\end{Definition}

The dual problem associated to the MK optimal transportation
problem is defined by :
 \b*
 D^0_2(\mu,\nu)
 :=
 \inf_{(\varphi,\psi)\in{\cal D}^0_2}
 \big\{ \mu(\varphi)+\nu(\psi) \big\},
 \e*
where, denoting $\varphi\oplus\psi(x,y):=\varphi(x)+\psi(y)$:
 \b*
 {\cal D}^0_2
 &:=&
 \big\{(\varphi,\psi):
       \varphi^+\in\L^1(\mu),\psi^+\in\L^1(\nu)
       ~\mbox{and}~
       \varphi\oplus\psi \ge c
 \big\}.
 \e*
and with $\mu(\varphi):=\int\varphi d\mu$, $\nu(\psi):=\int\psi d\nu$.

The dual problem $D^0_2(\mu,\nu)$ is the cheapest superhedging
strategy of the derivative security $c(X,Y)$ using the market
instruments consisting of $T-$maturity European calls and puts with
all possible strikes. The weak duality inequality
 \b*
 P^0_2(\mu,\nu)
 &\le&
 D^0_2(\mu,\nu)
 \e*
is immediate. For an upper semicontinuous payoff function $c$,
equality holds and an optimal probability measure $\P^*$  for the MK
problem $P^0_2$ exists, see e.g. Villani \cite{vil}.

Our main interest of this paper is the following one-dimensional
version of a result established by Brenier \cite{Brenier}, which
provides an interesting characterization of $\P^*$ in terms of the
so-called Fr\'echet-Hoeffding pushing forward $\mu$ to $\nu$,
defined by the map
 \be
 T_*
 :=
 F_\nu^{-1}\circ F_\mu,
 \ee
where $F_\nu^{-1}$ is the right-continuous inverse of $F_\nu$:
 \b*
 F_\nu^{-1}(t)
 :=
 \inf\{y:F_\nu(y)>x\}.
 \e*
In particular, the following result relates the MK optimal
transportation problem $P^0_2$ to the original Monge mass
transportation problem for a remarkable class of couplings $c$. This
result is more general, in particular the set of measures $\P_T$
induced by a map $T$ pushing forward $\mu$ to $\nu$ is dense in
${\cal P}_{\R^2}$ whenever $\mu$ is atomless and we consider compact
subsects of $\R^2$. For the purpose of our financial interpretation,
this result characterizes the structure of the worst case financial
market that the derivative security hedger may face, and
characterizes the optimal hedging strategies by the functions
$\varphi_*$ and $\psi_*$ defined up to an irrelevant constant by
 \begin{eqnarray}\label{integrability0}
 \varphi_*(x):=c\big(x,T_*(x)\big)-\psi_*\circ T_*(x),
 &\psi_*'(y):=c_y\big(T_*^{-1}(y),y\big),&
 x,y\in\R.
 \end{eqnarray}

\begin{Theorem}\label{thm-Frechet}{\rm (see e.g. \cite{vil}, Theorem 2.44)}
Let $c$ be upper semicontinuous with linear growth. Assume that the
partial derivative $c_{xy}$ exists and satisfies the Spence-Mirrlees
condition $c_{xy} > 0$. Assume further that $\mu$ has no atoms,
$\varphi^+_*\in\L^1(\mu)$ and $\psi^+_*\in\L^1(\nu)$. Then
\\
{\rm (i)} $P^0_2(\mu,\nu)=D^0_2(\mu,\nu)=\int c\big(x,T_*(x)\big)
\mu(dx)$,
\\
{\rm (ii)} $(\varphi_*,\psi_*)\in {\cal D}^0_2$, and is a solution
of the dual problem $D^0_2$,
\\
{\rm (iii)} $\P_*(dx,dy):=\mu(dx)\delta_{T_*(x)}(dy)$ is a solution
of the MK optimal transportation problem $P^0_2$, and is the unique
optimal transference map.
\end{Theorem}

\proof
We provide the proof for completeness, as our main result in this paper will be an adaptation
of the subsequent argument.
First, it is clear that $\P_*\in\Pc(\mu,\nu)$. Then $\E^{\P_*}[c(X,Y)]\le
P^0_2(\mu,\nu)$. We now prove that
 \begin{eqnarray}\label{phi*psi*OT}
 (\varphi_*,\psi_*)\in\Dc^0_2
 &\mbox{and}&
 \mu(\varphi_*)+\nu(\psi_*)=\E^{\P_*}[c(X,Y)].
 \end{eqnarray}
In view of the weak duality $P^0_2(\mu,\nu)\le D^0_2(\mu,\nu)$, this would imply that
$P^0_2(\mu,\nu)=D^0_2(\mu,\nu)$ and that $\P_*$ and $(\varphi_*,\psi_*)$ are solutions of
 $P^0_2(\mu,\nu)$ and $D^0_2(\mu,\nu)$, respectively.

Under our assumption that $\varphi_*\in\L^1(\mu)$, $\psi_*^+\in\L^1(\nu)$, notice that (\ref{phi*psi*OT}) is equivalent to:
 \b*
 0
 =
 H^0\big(x,T_*(x)\big)
 =
 \min_{y\in\R}
 H^0(x,y),
 &\mbox{where}&
 H^0:=\varphi_*\oplus\psi_*-c.
 \e*
The first-order condition for the last minimization problem provides the expression of $\psi'_*$ in (\ref{integrability0}), and the expression of $\varphi_*$ follows from the first equality. Since
 \b*
 H^0_y(x,y)
 &=&
 c_y\big(T_*^{-1}(y),y\big)-c_y(x,y)
 \;=\;
 \int_x^{T_*^{-1}(y)}c_{xy}(\xi,y)d\xi,
 \e*
it follows from the Spence-Mirrlees condition that $T_*(x)$ is the
unique solution of the first-order condition. Finally, we compute
that $H^0_{yy}\big(x,T_*(x)\big)=c_{xy}\big(x,T_*(x)\big)/T_*'(x)>0$
by the Spence-Mirrlees condition, where the derivatives are in the
sense of distributions. Hence $T_*(x)$ is the unique global
minimizer of $H(x,.)$. \ep

\vspace{5mm}

We observe that we may also formulate sufficient conditions on
the coupling $c$ so as to guarantee that the integrability
conditions $\varphi^+_*\in\L^1(\mu), \psi^+_*\in\L^1(\nu)$ hold
true. See \cite{vil}, Theorem 2.44.

\begin{Remark}[Mirror coupling: anti-monotone rearrangement map]
{\rm (i)} Suppose that the coupling function $c$ satisfies
$c_{xy}<0$. Then, the upper bound $P^0_2(\mu,\nu)$ is attained by the anti-monotone rearrangement map
 \b*
 \overline{\P}_*(dx,dy)
 :=
 \mu(dx)\delta_{\{\overline{T}_*(x)\}}(dy),
 &\mbox{where}&
 \overline{T}_*(x)
 :=
 F_\nu^{-1}\circ\big( 1- F_\mu(-x)\big).
 \e*
To see this, it suffices to rewrite the optimal transportation
problem equivalently with modified inputs:
 \b*
 \overline{c}(x,y):=c(-x,y),
 &\overline{\mu}(x):=\mu\big((-x,\infty)\big),&
 \overline{\nu}:=\nu,
 \e*
so that $\overline{c}$ satisfies the Spence-Mirrlees condition
$\overline{c}_{xy}>0$.
\\
{\rm (ii)} Under the Spence-Mirrlees condition $c_{xy}>0$, the lower
bound problem is explicitly solved by the anti-monotone
rearrangement. Indeed, it follows from the first part (i) of the
present remark that:
 $$
 \inf_{\P\in{\cal P}_2(\mu,\nu)}
 \!\E^\P\big[c(X,Y)\big]
 =
 -\sup_{\P\in{\cal P}_2(\mu,\nu)}
 \!\E^\P\big[-c(X,Y)\big]
 =
 -\E^{\overline{\P}_*}\big[-c(X,Y)\big]
 =
 \!\int\!\! c\big(x,\overline{T}_*(x)\big)\mu(dx).
 $$
\end{Remark}

\begin{Remark}
The Spence-Mirrlees condition is a natural requirement in the
optimal transportation setting in the following sense. The
optimization problem is not affected by the modification of the
coupling function from $c$ to $\bar c:=c+a\oplus b$ for
any $a\in\L^1(\mu)$ and $b\in\L^1(\nu)$. Since $c_{xy}=\bar
c_{xy}$, it follows that the Spence-Mirrlees condition is stable for
the above transformation of the coupling function.
\end{Remark}

\begin{Example}[Basket option] Let $c(x,y)=(x+y-k)^+$, for some $k\in\R$ (see
\cite{Daspremont-Elghaoui,lau} for multi-asset basket options). The
result of Theorem \ref{thm-Frechet} applies to this example as well,
as it is shown in \cite{vil} Chapter 2 that the
 regularity condition $c \in C^{1,1}$ is not needed. The upper bound is attained by the Fr\'echet-Hoeffding transference map $T_*:=F_\nu^{-1}\circ F_\mu$, and the optimal hedging strategy is:
 \b*
 \psi_*(y)
 =
 (y-\bar{y})^+,
 &\varphi_*(x)=\big(T_*(x)+x-k\big)^+
               - \big(T_*(x)-\bar{y}\big)^+ ,
 \e*
where $\bar y$ is defined by $T_*(k-\bar{y})=\bar{y}$.
\end{Example}

\subsection{The multi-marginals optimal transportation problem}

The previous results have been extended to the $n-$marginals optimal
transportation problem by Gangbo and
\'{S}wi\c{e}ch \cite{gan}, Carlier \cite{car}, and Pass \cite{Pass-fini}. Let $X=(X_1,\ldots,X_n)$ be a random
variable with values in $\R^n$, representing the prices at some
fixed time horizon of $n$ financial assets, and consider some upper
semicontinuous payoff function $c:\R^n\longrightarrow\R$ with linear
growth.

Let $\mu_1,\ldots,\mu_n\in\cal P_\R$ be the corresponding
marginal distributions, and $\mu:=(\mu_1,\ldots,\mu_n)$. The upper bound market price on the derivative security with a payoff function $c$ is defined by the optimal transportation problem:
 \be\label{P0n}
 P^0_n(\mu)
 :=
 \sup_{\P\in{\cal P}_n(\mu)}
 \E^\P\big[c(X)\big],
 &\mbox{where}&
 {\cal P}_n(\mu)
 :=\big\{\P\in{\cal P}_{\R^n}: X_i\sim_\P\mu_i, 1\le i\le n
   \big\}.
 \ee
Then, under convenient conditions on the coupling function $c$ (see Pass \cite{Pass-fini} for the most general ones), there exists a solution $\P_*$ to the MK optimal transportation problem $P^0_n(\mu)$ which is the unique optimal transference map defined by $T_*^i$, $i=2,\ldots,n$:
 \b*
 \P^*(dx_1, \ldots,dx_n)
 =
 \mu_1(dx_1)\prod_{i=2}^n \delta_{T^i_*(x_1)}(dx_i),
 &\mbox{where}&
 T^i_*=F_{\mu_i}^{-1}\circ F_{\mu_1},
 ~~i=2,\ldots, n.
 \e*
The optimal upper bound is then given by
 \b*
 P^0_n(\mu)
 =
 \int c\big(\xi,T^2_*(\xi),\ldots, T^n_*(\xi)\big)
      \mu_1(d\xi).
 \e*

\section{The Two-Marginals Martingale Transport Problem: Main Results}
\label{sect:mainresults}

The main objective of this paper is to obtain a version of the
Brenier theorem for the martingale transportation problem introduced
by Beiglb\"ock, Henry-Labord\`ere and Penkner \cite{BeiglbockHenry-LaborderePenkner} and
Galichon, Henry-Labord\`ere and Touzi \cite{GalichonHenry-LabordereTouzi}. A result in this
direction was first obtained by Hobson and Neuberger \cite{HobsonNeuberger} and
by Beiglb\"ock and Juillet \cite{BeiglbockJuillet}. In contrast with
the last reference, our result is an explicit extension of the
Fr\'echet-Hoeffding optimal coupling. We outline in Sections
\ref{sect-beiglbock} and \ref{sect-hobson} the main differences with
\cite{BeiglbockJuillet,HobsonNeuberger}.

\subsection{Problem formulation}

In the context of the financial motivation of Subsection
\ref{sect-OT}, we interpret the pair of random variables $X,Y$ as
the prices of the same financial asset at dates $t_1$ and $t_2$,
respectively, with $t_1<t_2$. Then, the no-arbitrage condition
states that the price process of the tradable asset is a martingale
under the pricing and hedging probability measure. We therefore
restrict the set of probability measures to:
 \b*
 \Mc_2(\mu,\nu)
 &:=&
 \big\{\P\in\Pc_2(\mu,\nu): \E^\P[Y|X]=X
 \big\}.
 \e*
where $\mu,\nu$ have finite first moment as in \eqref{firstmoment}.
This set of probability measures is clearly convex, and the
martingale condition implies that $\ell^{\nu}\le\ell^{\mu}\le
r^{\mu}\le r^{\nu}$. Throughout this paper, we shall denote
 \b*
 \delta F := F_\nu-F_\mu.
 \e*
By a classical result of Strassen \cite{Strassen}, $\Mc_2(\mu,\nu)$
is non-empty if and only if $\mu \preceq \nu$ in sense of convex ordering, i.e.
\begin{enumerate}
\item[(i)] $\mu,\nu$ have the same mean: $\int \xi d\delta F(\xi)=0$,
\item[(ii)] and $\int (\xi-k)^+\mu(d\xi) \leq
\int(\xi-k)^+\nu(d\xi)$, for all $k \in \R$. This condition can also be expressed as:
 \begin{eqnarray}\label{mulenu}
 \int_{[k,\infty)}\delta F(\xi)d\xi\le 0
 &\mbox{or, equivalently,}&
 \int_{[-\infty,k)}\delta F(\xi)d\xi\ge 0,
 ~~\mbox{for all}~~k\in\R,
 \end{eqnarray}
where the last equivalence follows from the first property (i).
\end{enumerate}
For completeness, we provide in Section \ref{sect-expMc} some
examples of probability measures in $\Mc_2(\mu,\nu)$ which are
commonly using by practitioners in quantitative finance.

Let $c:\R^2\longrightarrow\R$ be an upper semicontinuous function
with linear growth (or the condition (\ref{othercond})),
representing the payoff of a derivative security. In the present
context, the model-independent upper bound for the price of the
claim can be formulated as the following martingale optimal
transportation problem:
 \be
 P_2(\mu,\nu)
 &:=&
 \sup_{\P\in\Mc_2(\mu,\nu)}
 \E^\P\big[c(X,Y)\big],
 \ee

\begin{Remark}
When $\mu$ and $\nu$ have finite second moment, notice that
$\E^\P[(X-Y)^2]=-\E^\P[X^2]+\E^\P[Y^2]=\int \xi^2d\delta F(\xi)$ for
all $\P\in\Mc(\mu,\nu)$. Then, the quadratic case, which is the
typical example of coupling in the optimal transportation theory, is
irrelevant in the present martingale version.
\end{Remark}

We finally report the Kantorovich dual in the present martingale transport problem. Because of the possibility of
dynamic trading the financial asset between times $t_1$ and $t_2$, the set of dual variables is defined by:
 \be\label{defD}
 \Dc_2
 &:=&
 \big\{(\varphi,\psi,h):
       \varphi^+\in\L^1(\mu),\psi^+\in\L^1(\nu), h\in\L^0,
       ~\mbox{and}~
       \varphi\oplus\psi +h^\otimes\ge c
 \big\},
 \ee
where $\varphi\oplus\psi(x,y):=\varphi(x)+\psi(y)$, and
$h^\otimes(x,y):=h(x)(y-x)$. The dual problem is:
 \be
 D_2(\mu,\nu)
 &:=&
 \inf_{(\varphi,\psi,h)\in\Dc_2}
 \big\{\mu(\varphi)+\nu(\psi)\big\},
 \ee
and can be interpreted as the cheapest superhedging strategy of the derivative $c(X,Y)$ by dynamic trading on the
 underlying asset, and static trading on the European options with maturities $t_1$ and $t_2$. Since $c$ has
  linear growth and $\mu,\nu$ have finite first-order moments, the weak duality inequality:
 \be\label{weakduality}
 P_2(\mu,\nu)
 &\le&
 D_2(\mu,\nu)
 \ee
follows immediately from the definition of both problems. Under
suitable conditions on $c$, \cite{BeiglbockHenry-LaborderePenkner} proved the strong duality
result (i.e. equality holds), and showed the existence of a
maximizer $\P_*\in\Mc_2(\mu,\nu)$ for the martingale
transportation problem $P_2(\mu,\nu)$. However, existence does not
hold in general for the dual problem $D_2(\mu,\nu)$. An example of
non-existence is provided in \cite{BeiglbockHenry-LaborderePenkner}.

\subsection{Preliminaries}

Our objective in this section is to provide explicitly the
left-monotone martingale transport plan, as introduced by
Beiglb\"{o}ck and Juillet \cite{BeiglbockJuillet}.

\begin{Definition}\label{def-leftmonotone}
We say that $\P\in\Mc_2(\mu,\nu)$ is left-monotone (resp. right-monotone) if there exists a Borel set
$\Gamma\subset\R\times\R$ such that
$\P[(X,Y)\in\Gamma]=1,$ and for all $(x,y_1), (x,y_2),
(x',y')\in\Gamma$ with $x<x'$ (resp. $x>x'$), it must hold that
$y'\not\in(y_1,y_2)$.
\end{Definition}

Our main results hold for probability measures $\mu$, $\nu$
satisfying the following restriction.

\begin{Assumption}\label{assum-munu}
The probability measures $\mu$ and $\nu$ have finite first moments, $\mu\preceq\nu$ in convex order,
and $\mu$ has no atoms.
\end{Assumption}

Under this assumption, Theorem 1.5 and Corollary 1.6 of
\cite{BeiglbockJuillet} state that there exists a unique
left-monotone martingale transport plan $\P_*\in\Mc_2(\mu,\nu)$, and
that the graph of $\P_*$ is concentrated on two maps
$T_d,T_u:\R\longrightarrow\R$, $T_d(x)\le x\le T_u(x)$ for all
$x\in\R$, i.e. $\P_*[Y=T_d(X)]+\P_*[Y=T_u(X)]=1$.

\begin{Remark}\label{maximizerF}
The condition that $F_\mu$ is continuous in Assumption
\ref{assum-munu} implies that $\delta F$ is upper-semicontinuous, and therefore the local suprema of $\delta F$
are attained by maximizers.
\end{Remark}

For our construction, we introduce the functions:
 \begin{eqnarray}
 g(x,y)
 &:=&
 F_\nu^{-1}\big(F_\mu(x)+\delta F(y)\big),
 ~~x,y\in\R,
 \end{eqnarray}
where $F_\nu^{-1}$ denotes the right-continuous inverse of $F_\nu$, with $F_\nu^{-1}=\infty$ on $(1,\infty)$ and $F_\nu^{-1}=-\infty$ on $(-\infty,0)$. We also define for a measurable subset $A\in{\cal B}_\R$ such that $\delta F$ is increasing on $A$:
 \begin{equation}\label{GA}
 G^A(t,x)
 :=\!
 \int_{(-\infty,F_\nu^{-1}\circ F_\mu(x)]} \!\!\xi dF_\nu(\xi)
 -\!\int_{-\infty}^x \!\!\xi dF_\mu(\xi)
 +\!\int_{A\cap(-\infty,t]}  \!\!\big(g(x,\xi)-\xi\big)d\delta F(\xi),
 ~~t\le x \in\R.
 \end{equation}
In the last integral, notice that $g(x,\xi)-\xi\ge 0$, so that by the increase of $\delta F$ on $A$, the integral has a well-defined value in $(-\infty,\infty]$. It will be made clear in Section \ref{sect-construction} that these functions appear naturally when one imposes that $\P_*\in\Mc_2(\mu,\nu)$.

Notice that $G^A$ is right-continuous in $t$, and $G^A(-\infty,\infty)=0$, a consequence of the fact that $\mu$ and $\nu$ have the same mean. Our construction uses the following preliminary result, which needs the additional notation:
 \b*
 B_0:=\{x\in\R:~\delta F~\mbox{increasing to the right of}~x\},
 &x_0:=\inf B_0,&
 \e*
where we say that a function $\phi$ is increasing (resp. decreasing) to the right of $x$ if for all $\eps_0>0$, there exists $\eps\in(0,\eps_0)$ such that $\phi(x+\eps)>\phi(x)$ (resp. $\phi(x+\eps)<\phi(x)$).

Observe that $x_0=\infty$ if and only if $\mu=\nu$.

\begin{Lemma}\label{lem-stepi}
Assume $x_0<\infty$, let $m\in\R$ be a local maximizer of $\delta F$, and consider a Borel subset $A\subset (x_0,m]\cap B_0$. Denote $\bar A^m:=(x_0,m]\setminus A$, and assume that $\int_{\bar A^m}d\phi(\delta F)\ge 0$ for any non-decreasing function $\phi$. Then, there exists a unique scalar $t^A(x,m)$ such that, for all $x\ge m$ with $\delta F(x)\le\delta F(m)$,
 \b*
 t^A(x,m)\in A,
 &\mbox{and}&
 G^A\big(t^A(x,m)-,x\big)
   \le 0 \le
   G^A\big(t^A(x,m),x\big).
 \e*
Moreover, $\bar x(m):=\inf\{x>m:g\big(x,t^A(x,m)\big)\le x\}$ satisfies, whenever $\bar x(m)<\infty$,
 \be
 &\delta F\big(t^A(\bar x(m),m)\big)
 \le
 \delta F\big(\bar x(m)\big)
 \le
 \delta F\big(t^A(\bar x(m)-,m)\big)
 &
 \label{ineq-tA} \\
 &\mbox{and}~~
 \delta F
 ~~\mbox{is strictly increasing on a right neighborhood of}~~\bar x(m).&
 \nonumber
 \ee
\end{Lemma}

The proof of this lemma is reported in Subsection \ref{sect-construction}.

\subsection{Explicit construction}
\label{sect-explicitconstruction}

Our explicit construction requires an additional condition on $\delta F$.
Let $\mathbf{M}(\delta F)$ denote the collection of all points $m$ such that $\delta F$ is nondecreasing to the left of $m$, and decreasing to the right of $m$:
 \be\label{M(delta F)}
 \mathbf{M}(\delta F)
 &:=&
 \{m:\delta F'(m-)\le 0,~\mbox{and}~\delta F
           ~\mbox{decreasing to the right of}~m
   \big\},
 \ee
where we recall that $\delta F$ decreasing to the right of $m$ means that any right neighborhood of $m$ contains a point $m'$ such that $\delta F(m')<\delta F(m)$. Our general characterization of the left monotone transference plan will be obtained in Theorem \ref{thm-T*withaccumulation} as a limit of explicit monotone transference plans corresponding to an approximating sequence satisfying the following no right accumulation requirement.

\begin{Assumption}\label{assum-rightaccumulation}
$\mathbf{M}(\delta F)\cup\{x_0\}$ has no right accumulation point.
\end{Assumption}

\begin{Lemma}
Under Assumption \ref{assum-rightaccumulation}, the set $\mathbf{M}(\delta F)$ is countable.
\end{Lemma}

\proof
Under Assumption \ref{assum-rightaccumulation}, we have $\mathbf{M}(\delta F)=\cup_{n\in\N}\mathbf{M}_n(\delta F)$, where
 \b*
 \mathbf{M}_n(\delta F)
 :=\big\{m:\delta F
           ~\mbox{strictly decreasing on }
                    (m,m+\frac{1}{n}],
           ~\mbox{and}~\delta F'(m-)\le 0
   \big\}.
 \e*
Then the required result follows from the fact that $\mathbf{M}_n(\delta F)$ is countable.
\ep

\vspace{5mm}

We are now ready for the construction of the left-monotone
transference map $\P_*$. We first initialize the construction in
Step 0, and continue an iterative construction in the subsequent
steps.

\no \underline{\bf Step 0}:\quad If $x_0=-\infty$, we move to Step 1 of the construction. Otherwise, define:
 \begin{equation}\label{TdTu0}
 \begin{array}{c}
 T_d(x)
 \;=\;
 T_u(x)
 \;=\;
 x
 ~~\mbox{for}~~
 x\le x_0.
 \end{array}
 \end{equation}
If $x_0=\infty$, i.e. $\mu=\nu$, then this completes the construction of $(T_d,T_u)$. Otherwise, we continue with the following step.

\no \underline{\bf Step 1}:\quad By Assumption \ref{assum-rightaccumulation}, the function $\delta F$ increases
at the right of $x_0$. Consider the first point of decrease of
$\delta F$ (see Remark \ref{maximizerF}):
 \b*
 m_1
 :=
 \inf\big\{m>x_0:~\delta F~\mbox{decreasing on}~[m,m+\eps)
                          ~\mbox{for some}~\eps>0
     \big\},
 \e*
By the stochastic dominance $\mu\preceq\nu$, see (\ref{mulenu}), it follows that $\delta F$ is nondecreasing on $(x_0, m_1]$, $\delta F(m_1)>0$, and $\delta F$ is strictly increasing on the set
 \b*
 A_1:=(x_0,m_1] \cap B_0.
 \e*
\no We have  $\bar A^{m_1}_1=\emptyset$ and we are then in the
context of application of Lemma \ref{lem-stepi} with
$(m,A)=(m_1,A_1)$. Denoting $x_1
 := \bar x(m_1)$, we define the maps $T_d,T_u$ on $(-\infty,x_1)$:
 \begin{equation}\label{TdTu1}
 \begin{array}{c}
 T_d(x)
 \;=\;
 T_u(x)
 \;=\;
 x
 ~~\mbox{for}~~
 x_0<x\le m_1,
 \\
 T_d(x):=t^{A_1}(x,m_1),
 ~~T_u(x):=g\big(x,T_d(x)\big)
 ~~\mbox{for}~~
 m_1\le x< x_1.
 \end{array}
 \end{equation}
If $x_1=\infty$, this completes the construction. See Figure \ref{Ex1} below for such an example. Otherwise, it
follows from Lemma \ref{lem-stepi} that $\delta F$ is strictly
increasing at the right of $x_1$, whenever $x_1<\infty$. In this
case, we continue the construction denoting:
 \b*
 B_1:=B_0\setminus \big\{T_d\big([m_1,x_1)\big)
                         \cup\big[m_1,T_u(x_1)\big)\big\}
 =B_0\setminus\big(T_d(x_1),T_u(x_1)\big),
 \e*
where the last equality follows from the fact that $T_d$ is decreasing and $T_d(x) \leq x$, see Remark \ref{rem:Td}.

\no \underline{\bf Step 2}: The construction of this step falls in
the more general Step $i$ below, and is provided here for the
convenience of the reader.

Since $\mu\preceq\nu$ in (\ref{mulenu}), it follows that the set
of local maximizers after $x_1$ is not empty. Recall Assumption \ref{assum-rightaccumulation}, and let:
 \b*
 m_2
 &:=&
 \inf\big\{m\ge x_1:~\delta F~\mbox{decreasing on}~[m,m+\eps)
                          ~\mbox{for some}~\eps>0
     \big\},
 \\
 A_2
 &:=&
 (x_1,m_2]\cap B_1
 \;=\;
 (x_0,T_d(x_1))\cup(x_1,m_2],
 \e*
so that $\delta F$ is nondecreasing on $[x_1,m_2]$, and strictly increasing on $A_2$.
Moreover $\bar A_2^{m_2}=[T_d(x_1),x_1]$ and, since $\delta
F(T_d(x_1))\le\delta F(x_1)$ by \eqref{ineq-tA}, we see that $\int_{\bar A_2^{m_2}}d\phi(\delta F)\ge 0$ for all nondecreasing function $\phi$.

Then, we may apply Lemma \ref{lem-stepi} with $(m,A)=(m_2,A_2)$. Denoting $x_2 := \bar x(m_2)$, we may
define the maps $T_d,T_u$ on $[x_1,x_2)$:
 \begin{equation}\label{TdTu2}
 \begin{array}{c}
 T_d(x)
 \;=\;
 T_u(x)
 \;=\;
 x
 ~~\mbox{for}~~
 x_1<x\le m_2,
 \\
 T_d(x):=t^{A_2}(x,m_2),
 ~~T_u(x):=g\big(x,T_d(x)\big)
 ~~\mbox{for}~~
 m_2\le x< x_2.
 \end{array}
 \end{equation}
If $x_2=\infty$, this completes the construction. See Figure \ref{Ex2MaxLocaux} below for such an example. Otherwise, it
follows from Lemma \ref{lem-stepi} that $\delta F$ is strictly
increasing at the right of $x_2$, whenever $x_2<\infty$. In this
case, we continue the construction denoting:
 \b*
 B_2:=B_1\setminus \big\{T_d\big([m_2,x_2)\big)
                         \cup\big[m_2,T_u(x_2)\big)\big\}.
 \e*

\no \underline{\bf Step i}:\quad Suppose that $(T_d,T_u)$ are
defined on $(-\infty,x_i)$ for some $x_i$ with $\delta F$ strictly
increasing at the right of $x_i$, and let a subset
$B_i:=B_{i-1}\setminus\{T_d([m_i,x_i])\cup[m_i,x_i]\}\subset B_0$ be
given so that  by definition, we have
 \be \label{Td(xi)}
 G^{A_i}(T_d(x_i),x_i)
 &\geq&
 0.
 \ee
and $A_i$ is obtained iteratively from the previous steps as:
 \b*
 A_i
 =
 (x_0,m_i]
 \setminus
 \big[\cup_{j<i} \big\{T_d\big([m_j,x_j)\big)
                       \cup\big[m_j,T_u(x_j)\big)\big\}\big].
 \e*
\no Since $\mu\preceq\nu$ in (\ref{mulenu}), it follows
that the set of local maximizers after $x_i$ is not empty. Recall Assumption \ref{assum-rightaccumulation}, and let:
 \b*
 m_{i+1}
 :=
 \inf\big\{m\ge x_i: \delta F~\mbox{decreasing on}~[m,m+\eps)
                          ~\mbox{for some}~\eps>0
     \big\},
 \e*
and
 \b*
 A_{i+1}
 &:=&
 (x_0,m_{i+1}) \cap B_{i}
 \;=\;
 (x_0,m_{i+1}]
 \setminus
 \big[\cup_{j<i} \big\{T_d\big([m_j,x_j)\big)
                       \cup\big[m_j,T_u(x_j)\big)\big\}\big],
 \e*
so that $\delta F$ is strictly increasing on $A_{i+1}$.

We  observe that $T_d(x_i)\not\in
[T_d(x_j),T_u(x_j)]$ for any $j<i$, which expresses
that our construction provides the left-monotone martingale
transport plan, see Definition \ref{def-leftmonotone}.
Since $\delta F(T_d(x_i))\le\delta F(x_i)$ by \eqref{ineq-tA}, we have also that $\int_{A^{m_{i+1}}_{i+1}}d\phi(\delta F)\ge 0$ for all nondecreasing function $\phi$.

We have thus verified that the conditions of Lemma
\ref{lem-stepi} are satisfied by the pair $(m_{i+1},A_{i+1})$, and we may then define the maps $T_d,T_u$ on
$[x_i,x_{i+1})$ by:
 \begin{equation}\label{TdTui}
 \begin{array}{c}
 T_d(x)=T_u(x)=x
 ~~\mbox{for}~~
 x_i\le x\le m_{i+1},
 \\
 T_d(x):=t^{A_{i+1}}(x,m_{i+1}),
 ~~T_u(x):=g\big(x,T_d(x)\big)
 ~~\mbox{for}~~
 m_{i+1}\le x<x_{i+1}
 := \bar x(m_{i+1}).
 \end{array}
 \end{equation}
\no If $x_{i+1}=\infty$, the construction is complete. Otherwise, it
follows from Lemma \ref{lem-stepi} that $\delta F$ is strictly
increasing at the right of $x_{i+1}$, whenever $x_{i+1}<\infty$. In
this case, we also update:
 \b*
 B_{i+1}
 :=
 B_i\setminus \big\{T_d\big([m_{i+1},x_{i+1})\big)
                    \cup\big[m_{i+1},T_u(x_{i+1})\big)
              \big\},
 \e*
and we continue with an additional step.

\no \underline{\bf Case of accumulation}: \quad \quad It may happen
that the increasing sequence $(m_i)_i$ converges to some
$m^1<\infty$. Then, as the number of steps $i$ tends to infinity,
the above construction defines the maps $(T_d,T_u)$ on
$(-\infty,m^1)$.

In this case, under Assumption \ref{assum-rightaccumulation} which excludes any right accumulation of local maxima, we may start again the construction exactly as in Step i, with $m_{i+1}=m^1_1$. After possibly i steps, this defines
$(m^1_j,x^1_j)_{j\le i}$ which either meets the requirement
$x^1_i=\infty$, or accumulates. Recall that the set $\mathbf{M}(\delta)$ of \eqref{M(delta F)} is countable under our Assumption \ref{assum-rightaccumulation}. Since the set of possible accumulation points $m^k$ is a subset of $\mathbf{M}(\delta F)$, it is at most countable. Then, by transfinite induction,
 \b*
 \mbox{\it we relabel the sequence}
 &(m^k_j,x^k_j)_{j,k}&
 \mbox{\it as a new sequence that we rename}~~
 (m_i,x_i)_{i\ge 0}.
 \e*

\begin{Remark}[Some properties of $T_d$]\label{rem:Td} From the above construction of $T_d$, we see that
\\
{\rm (i)} $T_d$ is right-continuous. Moreover, on each interval $(m_i,x_i)$, it is non-increasing and flat if and only if it reaches an atom of $F_\nu$.
\\
{\rm (ii)} In general, the restriction of $T_d$ to $\cup_{i\ge 0}(m_i,x_i)$ fails to be non-decreasing. However, for $i\neq j$,
 we have $T_d\big((m_i,x_i)\big)\cap T_d\big((m_j,x_j)\big)=\emptyset$. Consequently, the right-continuous inverse
 $T_d^{-1}$ of $T_d$ is well defined.
\\
{\rm (iii)} Let $I=(a,b)\subset T_d([m_i,x_i])$ be such that $\delta F$ is flat on $I$, and  $\delta F$ increases at the
right of $b$ and at the left of $a$. Then, whenever $T_d$ reaches the right endpoint $b$, it jumps from $b$ to $a$,
i.e. $\Delta T_d\big(T_d^{-1}(b)\big)=a-b$.
\\
{\rm (iv)} Let $x$ be such that $T_d(x)=T_u(x)=x$. Then, $\{x'\neq x:T_u(x')=x\}=\emptyset$,
$\{x'\neq x:T_d(x')=x\}\neq\emptyset$, and reduces to a single point set if $\Delta F_\nu(x)=0$.
Otherwise, if $x$ is an atom of $F_\nu$, the last set has a positive measure under $F_\mu$.
\end{Remark}

\begin{Remark}[Some properties of $T_u$]\label{rem:Tu}
From the above construction of $T_u$, we see that
\\
{\rm (i)} $T_u([m_i,x_i])\subset [m_i,x_i]$, and $T_u(x)>x$ for $x\in(m_i,x_i)$ for all $i$.
\\
{\rm (ii)} $T_u$ is right continuous with discontinuity points $\{x:\Delta
F_\nu(T_d(x))>0\}$, recall that
$F_\mu$ is continuous.
\\
{\rm (iii)} $T_u$ is nondecreasing, and strictly increasing on the support of  $\mu$. The last property will be clear from Theorem \ref{thm-defT} (ii) below, and implies that the right-continuous inverse $T_u^{-1}$ of $T_u$ is well-defined.
\end{Remark}

\subsection{The left-monotone martingale transport plan}

\no The last construction provides our martingale version of the
Fr\'echet-Hoeffding coupling:
 \begin{equation}\label{T*}
 T_*(x,dy)
 :=
 \1_{D}(x)
 \delta_{\{x\}}(dy)
 +\1_{D^c}(x)
  \big[q(x)\delta_{\{T_u(x)\}}(dy)
                     +(1-q)(x)\delta_{\{T_d(x)\}}(dy)
            \big],
 \end{equation}
\no where $x_{-1}=-\infty$, $m_0:=x_0$,
 \be\label{D}
 D:=\cup_{i\ge 0}(x_{i-1},m_i]
 &\mbox{and}&
 q(x):=\frac{x-T_d(x)}{T_u(x)-T_d(x)}.
 \ee
\no Observe that $T_d(x)\le x\le T_u(x)$ from our previous construction.
Therefore, $q$ takes values in $[0,1]$.

\begin{Theorem}\label{thm-defT}
Let Assumptions \ref{assum-munu} and \ref{assum-rightaccumulation} hold true. Then,
\\
{\rm (i)} the probability measure $\P_*(dx,dy):=\mu(dx)T_*(x,dy)$
is the unique left-monotone transport plan in $\Mc_2(\mu,\nu)$;
\\
{\rm (ii)} moreover $T_u$ and $T_d$ solve the following ODEs:
 \b*
 d(\delta F\circ T_d)
 =
 -(1-q)dF_\mu,
 ~~
 d(F_\nu\circ T_u)
 =
 qdF_\mu
 &\mbox{whenever}&
 x\in[m_i,x_i)
 ~~\mbox{and}~~
 T_d(x)\in\mbox{\rm int}(A_i).
 \e*
\end{Theorem}

\no The proof is reported in Section \ref{sect-construction}. The next result characterizes the left-monotone transference map in the case where $\delta F$ does not satisfy Assumption \ref{assum-rightaccumulation}.

\begin{Theorem}\label{thm-T*withaccumulation}
Let Assumption \ref{assum-munu} hold true, and let $(\mu_n,\nu)_{n\ge 1}\subset\Pc_\R$ be such that $\mu_n\longrightarrow\mu$ and $\nu_n\longrightarrow\nu$, weakly, and $(\mu_n,\nu_n)$ satisfies Assumptions \ref{assum-munu} and \ref{assum-rightaccumulation}. For all $n\ge 1$, define the corresponding $T^n_*$ as in \eqref{T*}, and the corresponding $\P_*^n(dx,dy):=\mu_n(dx)T^n_*(x,dy)$.

Then $\P^n_*$ converges weakly towards the unique left-monotone transference map.
\end{Theorem}

\proof By following the proof of Proposition 2.4 of \cite{BeiglbockHenry-LaborderePenkner}, it follows from Lemma 4.4 p56 in Villani \cite{vil} that the sequence $(\P^n_*)_{n\ge 1}$ is weakly compact. Then, after possibly passing to subsequence, $\P^n_*\longrightarrow\hat\P$, weakly, for some $\hat\P\in\Mc(\mu,\nu)$. To prove the required result, we shall prove that $\hat\P$ is a left-monotone transference map; then, from the uniqueness result of Theorem 1.5 in \cite{BeiglbockJuillet}, we may deduce that $\hat\P$ does not depend on the chosen subsequence.

Assume to the contrary that $\hat\P$ is not left-monotone. Then there exists a support $\hat\Gamma$ of $\hat\P$ such that
 \be\label{y'inydyu}
 (x,y_d),~(x,y_u),~(x',y')\in\hat\Gamma~,
 ~y_d<y_u,~x'>x,
 &\mbox{and}&
 y'\in(y_d,y_u).
 \ee
To obtain the required contradiction, we prove below that there exist sequences $(x^n,y_d^n)_n$, $(x^n,y_u^n)_n$, $(x'_n,y'_n)_n$ in a support of $\P^n_*$ such that $(x^n,x'_n)\longrightarrow (x,x')$, and $(y_d^n,y_u^n,y'_n)\longrightarrow (y_d,y_u,y')$. By the left-monotonicity of $\P^n_*$ for all $n$, we have $y'_n\not\in(y_d^n,y_u^n)$, and we obtain by sending $n\to\infty$ that $y'\not\in(y_d,y_u)$, contradicting \eqref{y'inydyu}.

We finally prove that if $(x,y)\in\hat\Gamma$, then there exists a sequence $(x_n,y_n)$ and a support of $\P_n$ such that $(x_n,y_n)\longrightarrow (x,y)$. For an arbitrary $\eps>0$, let $\varphi$ a continuous function with support in $B_\eps(x,y)$, the open ball centered at $(x,y)$ with radius $\eps$. Then, it follows from the weak convergence of $\P^n_*$ towards $\hat\P$ that $\E^{\P^n_*}[\varphi(X,Y)]\longrightarrow\E^{\hat\P}[\varphi(X,Y)]$, and the required result follows from the arbitrariness of $\eps>0$.
\ep

\vspace{5mm}

We conclude this subsection by the following remarkable property of
$T_d$.

\begin{Proposition}\label{prop-Tdx*}
Let Assumptions \ref{assum-munu} and \ref{assum-rightaccumulation} hold true. Let $i\ge 1$ be such that $\delta F$ is not flat at the left of $m_i$. Then $T_d(m_i+)=m_i$. If in addition $F_\mu, F_\nu$ are twice differentiable near $m_i$, then:
 \b*
 T_d'(m_i+)=-1/2
 &\mbox{and}&
 T_d''(m_i)=+\infty.
 \e*
\end{Proposition}

\no {\bf Proof}
By construction, we have $T_d(m_i+)=m_i$. Denoting $\eps:=x-T_d(x)$,
$f_\mu:=F'_\mu$, $f_\nu:=F'_\nu$, $\delta f:=f_\nu-f_\mu$, and
recalling that $g(x,x)=x$, we see by direct calculation that
 \b*
 g(x,T_d)-x
 &\!\!=&\!\!
 -\eps\frac{\delta f}{f_\nu}(x)
 +\frac{\eps^2}{2}\Big(\frac{\delta f'}{f_\nu}
                    +\Big(\frac{\delta f}
                               {f_\nu}\Big)^2
                     \frac{f_\nu'}{f_\nu}
               \Big)(x)
 +\circ(\eps^2),
 \\
 \delta f\circ T_d(x)
 &\!\!=&\!\!
 -\eps\delta f'(x)
 +\circ(\eps).
 \e*
where $\circ$ is a continuous function with $\circ(0)=0$. Observe
that $\delta f>0$ near $m_i$ by the definition of $m_i$. Plugging
the above expansion in the ODE satisfied by $T_d$, we see that:
 \b*
 T_d'(x)
 &=&
 -\frac{\frac{\delta f}{f_\nu}
        +\frac12\eps\left(\frac{\delta f'}{f_\nu}
                          +\left(\frac{\delta f}{f_\nu}
                           \right)^2\frac{f_\nu'}{f_\nu}
                    \right)+\circ(\eps)}
       {1-\frac{\delta f}{f_\nu}
        +\frac12\eps\left(\frac{\delta f'}{f_\nu}
                           +\left(\frac{\delta f}{f_\nu}
                            \right)^2\frac{f_\nu'}{f_\nu}
                     \right)+\circ(\eps)}
  \frac{f_\mu}{\delta f-\eps\delta f'+\circ(\eps)}
  (x).
 \e*
\no We then take the limit as $x\searrow m_i$, so that $\eps\searrow
0$ and $\delta f(x)\longrightarrow 0$ by the definition of $m_i$.
This leads to $T_d'(x)\longrightarrow -1/2$.

Finally, we compute $T_d''(m_i)$. By the ODE satisfied by $T_d$
and the smoothness of $g$, it follows that $T_d'$ is differentiable
at any $x>m_i$. We then differentiate the ODE satisfied by $T_d$,
and use Taylor expansions. The result follows from direct
calculation by sending $x\searrow m_i$.
\ep

\subsection{Martingale version of the Brenier Theorem}

We next introduce a remarkable triple of dual variables
corresponding to a smooth coupling function $c$. Recall the set $D$
defined in (\ref{D}) on which we have $T_d(x)=T_u(x)=x$, $x\in D$,
and the right-continuous inverse functions $T_d^{-1}, T_u^{-1}$
defined in Remark \ref{rem:Td} (ii) and Remark \ref{rem:Tu} (iii).

The dynamic hedging component $h_*$ is defined up to a constant, on each continuity interval, by:
 \begin{equation}\label{defh}
 h_*'=\frac{c_x(.,T_u)-c_x(.,T_d)}{T_u-T_d}
 ~\mbox{on}~D^c,~
 h_*=h_*\circ T_d^{-1}+c_y(.,.)-c_y(T_d^{-1},.)
 ~\mbox{on}~D.
 \end{equation}
The payoff function $\psi_*$ is defined up to a constant on each continuity interval by:
 \begin{eqnarray}\label{defg}
 \psi_*'=c_y(T_u^{-1},.)-h_*\circ T_u^{-1}
 &\mbox{on}~~D^c,&
 \psi_*'=c_y(T_d^{-1},.)-h_*\circ T_d^{-1}
 ~~\mbox{on}~~D.
 \end{eqnarray}
The corresponding function $\varphi_*$ is given by:
 \begin{eqnarray}\label{deff}
 \varphi_*(x)
 &=&
 \E^{\P_*}\big[c(X,Y)-\psi_*(Y)|X=x\big]
 \\
 &=&
 q(x)\big(c(x,.)-\psi_*\big)\circ T_u(x)
 +\big(1-q(x)\big)\big(c(x,.)-\psi_*\big)\circ T_d(x),
 ~~
 x\in\R.
 \nonumber
 \end{eqnarray}
Finally, we define $h_*$ and $\psi_*$ from \eqref{defh}-\eqref{defg} by imposing that
 \be\label{deltaHcontinuous}
 \mbox{the function}
 &c(.,T_u)-\psi_*(T_u)
  -[c(.,T_d)-\psi_*(T_d)]
  -(T_u-T_d)h&
  \mbox{is continuous.}
 \ee

\begin{Theorem}\label{thm-Brenier}
Let $\mu,\nu$ be as in Assumptions \ref{assum-munu} and \ref{assum-rightaccumulation}. Assume further that $\varphi_*^+\in\L^1(\mu)$,
$\psi_*^+\in\L^1(\nu)$, and that the partial derivative of the coupling function $c_{xyy}$ exists and
$c_{xyy}>0$ on $\R\times\R$. Then:
\\
{\rm (i)} $(\varphi_*,\psi_*,h_*)\in\Dc_2$,
\\
{\rm (ii)} the strong duality holds for the martingale
transportation problem, $\P_*$ is a solution of $P_2(\mu,\nu)$, and
$(\varphi_*,\psi_*,h_*)$ is a solution of $D_2(\mu,\nu)$:
 $$
 \int c\big(x,T_*(x,dy)\big)\mu(dx)
 \;=\;
 \E^{\P_*}\big[c(X,Y)]
 \;=\;
 P_2(\mu,\nu)
 \;=\;
 D_2(\mu,\nu)
 \;=\;
 \mu(\varphi_*)+\nu(\psi_*).
 $$
\end{Theorem}

\begin{Remark}[Mirror coupling: the right-monotone martingale transport plan] \label{rem-mirror}~~
\\
{\rm (i)} Suppose that $c_{xyy}<0$. Then, the upper bound
$P_2(\mu,\nu)$ is attained by the right-monotone martingale transport map
 \b*
 \bar\P_*(dx,dy)
 :=
 \bar\mu(dx)\bar T_*(x,dy),
 \e*
where $\bar T_*$ is defined as in (\ref{T*}) with the pair of probability measures $(\bar\mu,\bar\nu)$:
 \b*
 F_{\bar\mu}(x):=1-F_\mu(-x),
 &\mbox{and}&
 F_{\bar\nu}(y)
 :=
 1-F_\nu(-y).
 \e*
To see this, we rewrite the optimal transportation
problem equivalently with modified inputs:
 \b*
 \bar c(x,y):=c(-x,-y),
 &\bar\mu\big((-\infty,x]\big):=\mu\big([-x,\infty)\big),&
 \bar\nu\big((-\infty,y]\big)
 :=\nu\big([-y,\infty)\big),
 \e*
so that $\bar c_{xyy}>0$, as required in Theorem
\ref{thm-Brenier}. Note that the martingale constraint is preserved by the map $(x,y) \rightarrow (-x,-y)$.
\\
{\rm (ii)} Suppose that $c_{xyy}>0$. Then, the lower bound problem is explicitly solved by the right-monotone
 martingale transport plan. Indeed, it follows from the first part (i) of the present remark that:
 $$
 \inf_{\P\in{\cal M}_2(\mu,\nu)}
 \!\!\E^\P\big[c(X,Y)\big]
 =
 -\sup_{\P\in{\cal M}_2(\mu,\nu)}
 \!\!\E^\P\big[-c(X,Y)\big]
 =
 \E^{\bar\P_*}\big[c(X,Y)\big]
 =
 \int \!\!c\big(x,\bar T_*(x,dy)\big)\mu(dx).
 $$
\end{Remark}

\begin{Remark}\label{symmetrytransformation}
The martingale counterpart of the Spence-Mirrlees condition is
$c_{xyy}> 0$. We now argue that this condition is the natural
requirement in the present setting. Indeed, the optimization problem
is not affected by the modification of the coupling function from
$c$ to $\bar c(x,y):=c(x,y)+a(x)+b(y)+h(x)(y-x)$ for any
$a\in\L^1(\mu)$, $b\in\L^1(\nu)$, and $h\in\L^0$. Since
$c_{xyy}=\bar c_{xyy}$, it follows that the condition $c_{xyy}> 0$
is stable for the above transformation of the coupling function.
\end{Remark}

\subsection{Comparison with Beiglb\"ock and Juillet \cite{BeiglbockJuillet}}
\label{sect-beiglbock}

\begin{figure}[tbp]
\begin{center}
\includegraphics[width=7cm,height=7cm]{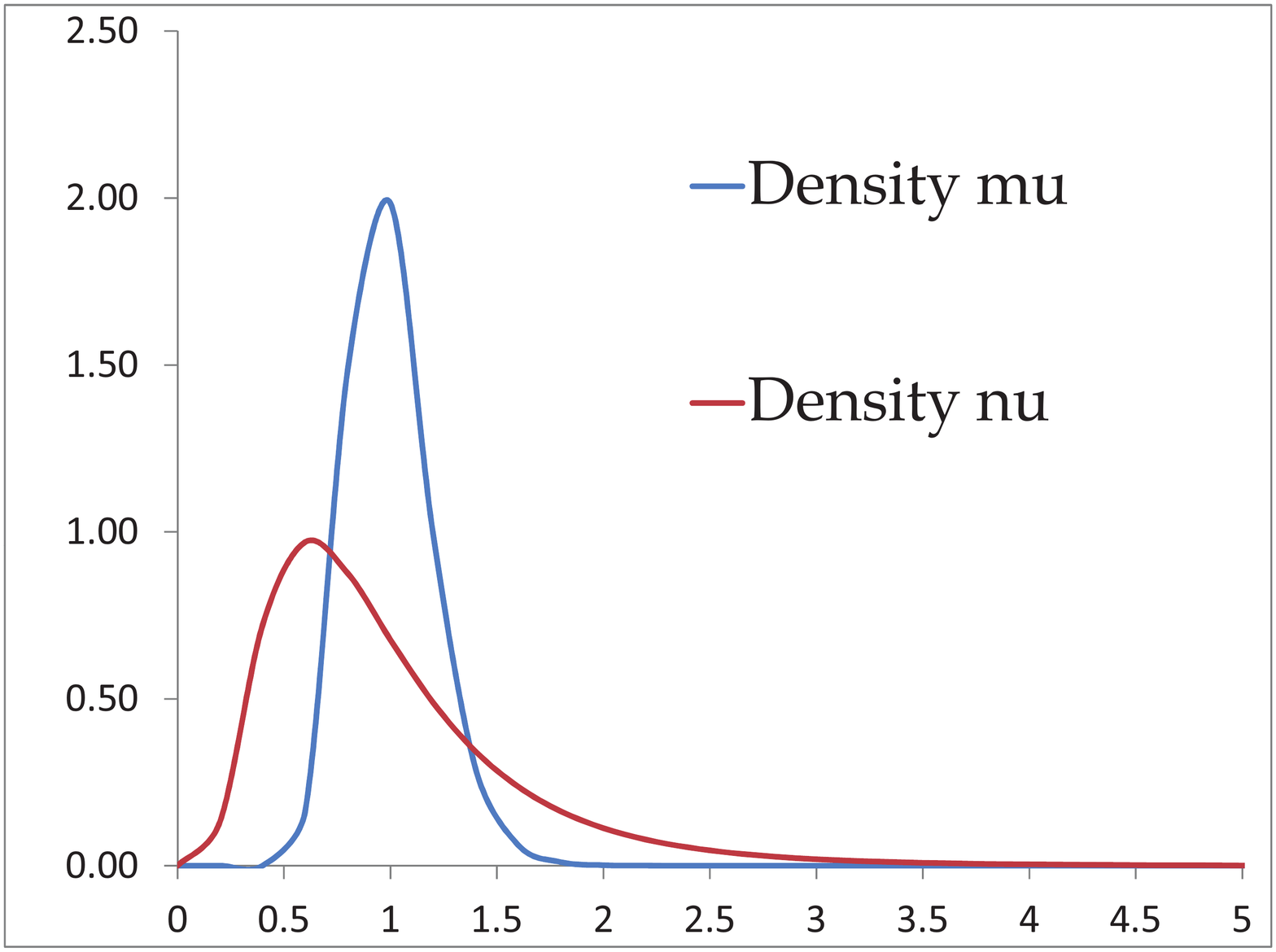}
\includegraphics[width=7cm,height=7cm]{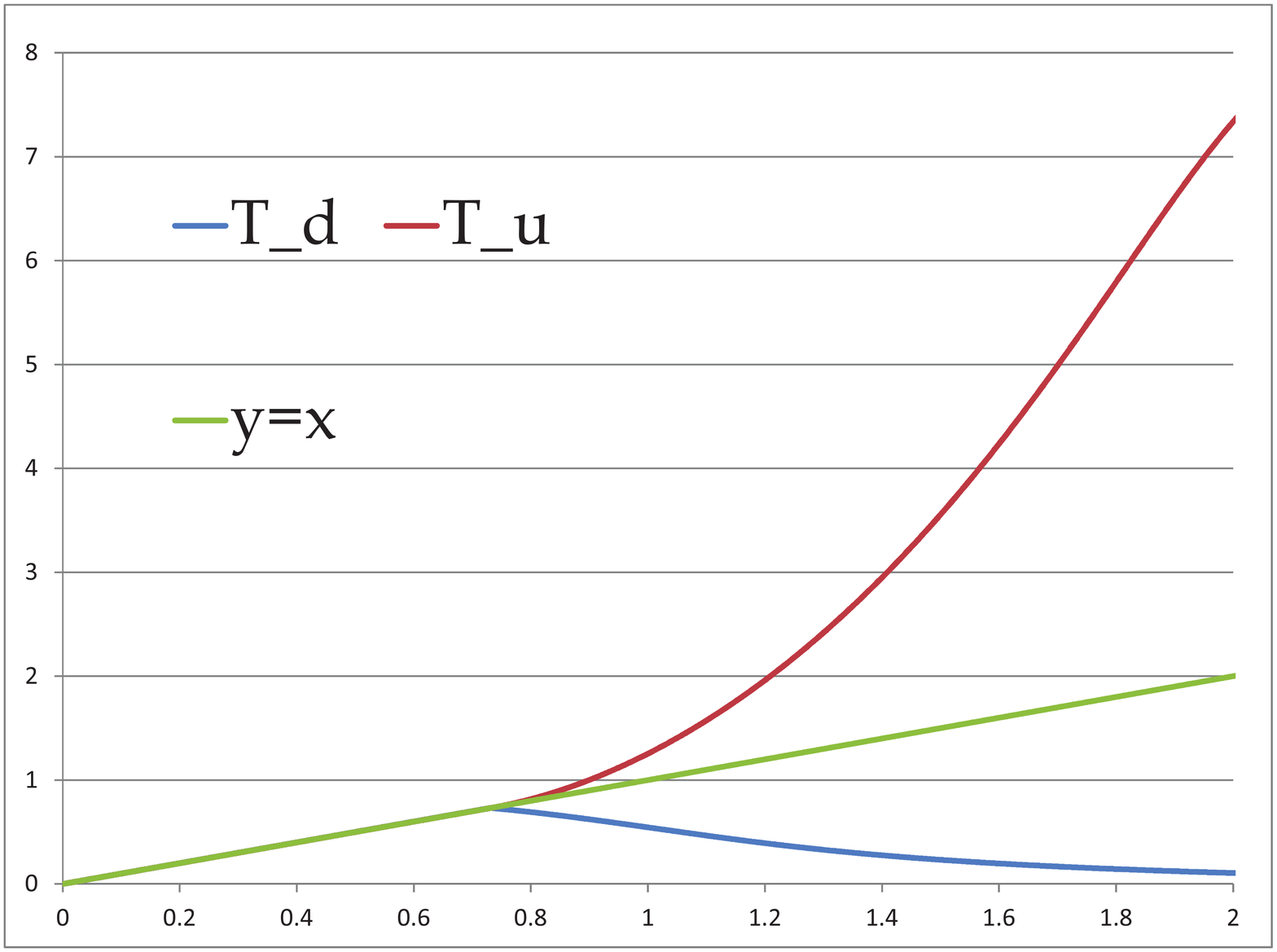}
\end{center}
\caption{Maps $T_d$ and $T_u$ built from two log-normal densities
with variances $0.04$ and $0.32$. $m_1=0.731$.} \label{Ex1}
\end{figure}

The notion of left-monotone martingale transport was introduced by Beiglb\"{o}ck and Juillet \cite{BeiglbockJuillet},
 with an existence and uniqueness result, see Theorem 1.7 and Theorem 6.2.
\begin{enumerate}
\item We first show that their conditions on the coupling function fall in
the context of our Theorem \ref{thm-Brenier}:
\begin{itemize}
\item The first class of couplings considered in \cite{BeiglbockJuillet} is of the form $c(x,y)=h(y-x)$
for some differentiable function $h$ whose derivative is strictly
concave. Notice that this form of coupling essentially falls under our condition $c_{xyy}>0$.
\item The second class of couplings considered in \cite{BeiglbockJuillet} is of the form $c(x,y)=\psi(x)\phi(y)$
where $\psi$ is a non-negative decreasing function and $\phi$ a non-negative strict concave function. This class
also essentially falls under our condition that $c_{xyy}>0$.
\end{itemize}
\item The proof of \cite{BeiglbockJuillet} does not use the dual
formulation of the martingale optimal transport problem. They rather
extend the concept of cyclical monotonicity to the martingale
context. As a consequence, \cite{BeiglbockJuillet} only provides an
existence result and does not contain any explicit characterization
of the maps $(T_d,T_u)$ and the optimal semi-static hedging strategy
$(\varphi_*,\psi_*,h_*)$.
\item Our left-monotone martingale transport map $T_*$ coincides with the {\it left-curtain coupling} whose
existence (and uniqueness) is stated in Theorem 4.18 of \cite{BeiglbockJuillet}.
\item Our construction agrees with the example of two Log-normal distributions $\mu_0=e^{\Nc(-\sigma^2_1/2,\sigma^2_1)}$ and
$\nu_0=e^{\Nc(-\sigma^2_2/2,\sigma^2_2)}$, $\sigma^2_1<\sigma_2^2$, illustrated in
Figure 2 of \cite{BeiglbockJuillet}. By using our construction, we reproduce the left-monotone transference map in Figure \ref{Ex1}.
Indeed, in this case, $x_0=-\infty$, $\delta F$ has a unique local
maximizer $m_1$, which is then the global maximizer of $\delta F$,
and $x_1=\infty$. The left-monotone transport plan is explicitly
obtained from our construction after Step 1, i.e. no further steps
are needed in this case.
\end{enumerate}

\begin{Example}
We provide an example where $\delta F$ has two local maxima and the construction needs two steps. Let $\mu$ and $\nu$ be defined by
 \b*
 \mu_1=\Nc(1,0.5)
 &\mbox{and}&
 \nu_1(x)={1 \over 3} \big[
      \Nc(1,2)+\Nc(0.6,0.1)+\Nc(1.4,0.3) \big].
 \e*
Clearly $\mu$ and $\nu$ have mean $1$, and $\mu\preceq\nu$. We also immediately check that $\delta F$ has two local maxima $m_1=-0.15$ and $m_2=0.72$. Figure \ref{Ex2MaxLocaux} below reports the maps $T_u$ and $T_d$ as obtained from our construction.

\begin{figure}[tbp]
\begin{center}
\includegraphics[width=8cm,height=9cm]{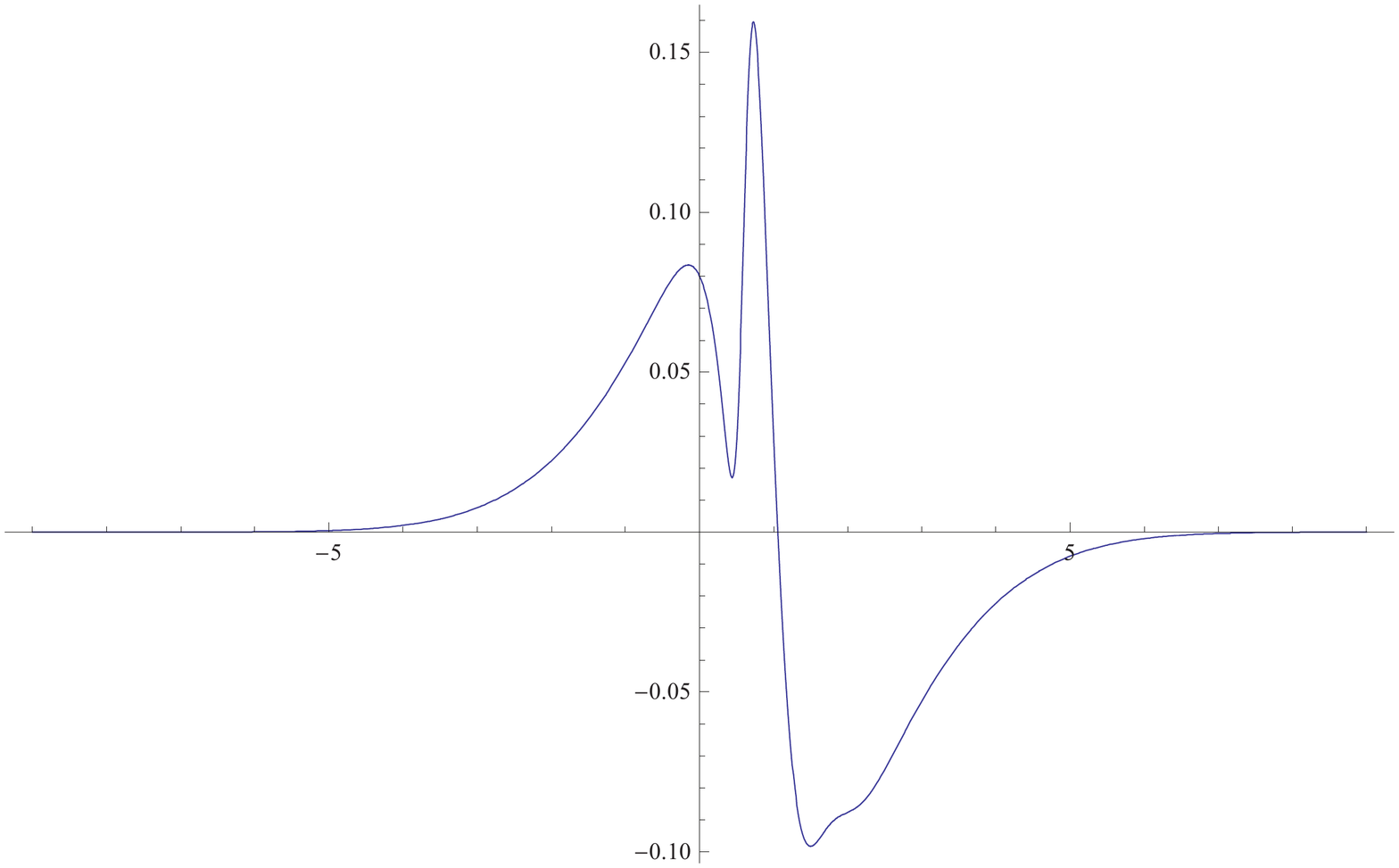}
\includegraphics[width=8cm,height=9cm]{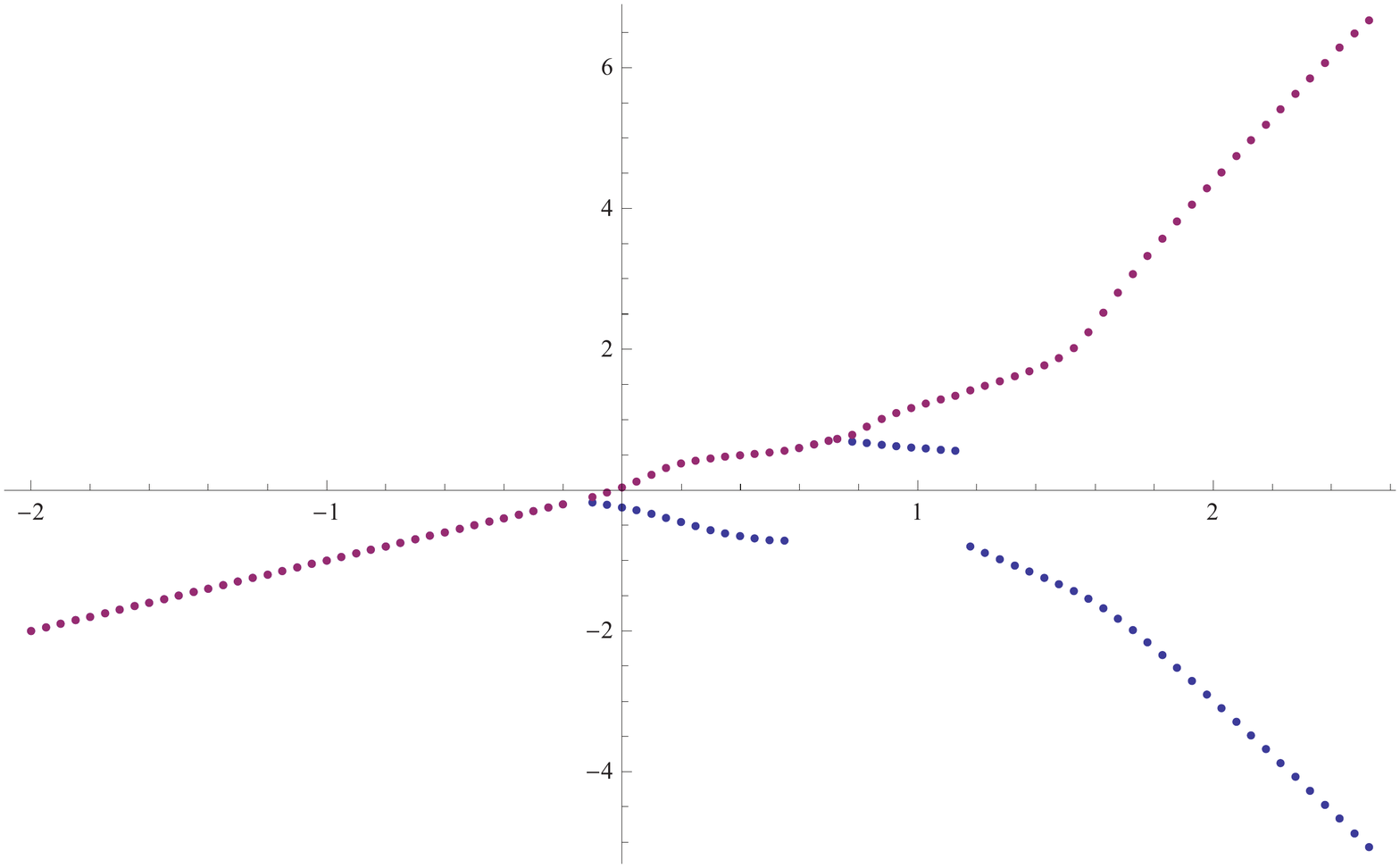}
\end{center}
\caption{$\delta F$ has two local maxima (left), and $T_d,T_u$
corresponding to $\mu_1,\nu_1$ (right).} \label{Ex2MaxLocaux}
\end{figure}

\end{Example}

\subsection{Comparison with Hobson and Neuberger \cite{HobsonNeuberger}}
\label{sect-hobson}

Our Theorem \ref{thm-Brenier} does not apply to the coupling function $c(x,y)=|x-y|$ considered by Hobson and
Neuberger \cite{HobsonNeuberger}. More importantly, the corresponding maps $T^{\mbox{\sc hn}}_u$ and $T^{\mbox{\sc hn}}_d$
introduced in \cite{HobsonNeuberger} are both nondecreasing with $T^{\mbox{\sc hn}}_d(x)<x<T^{\mbox{\sc hn}}_u(x)$ for all
$x\in\R$. So our solution $(T_d,T_u)$ is of a different nature and in contrast with the above
$(T^{\mbox{\sc hn}}_d,T^{\mbox{\sc hn}}_d)$, our left-monotone martingale transport map $T_*$ does not
depend on the nature of the
coupling function $c$ as long as $c_{xyy}>0$.

\no However, by following the line of argument of the proof of Theorem \ref{thm-Brenier}, we may recover the solution
of Hobson and Neuberger \cite{HobsonNeuberger}. As a matter of fact, our method of proof is similar to that of \cite{HobsonNeuberger}, as
the dual problem $D_2$ is exactly the Lagrangian obtained by the penalization of the objective function by Lagrange
multipliers.

\subsection{Some examples}

\begin{Example}[Variance swap]\label{ex-Varswap} The coupling in this case is $c(x,y)= \ln^2 \left({y \over x} \right)$ where $\mu$ and $\nu$ have support in $(0,\infty)$. In particular, it satisfies the requirement of Theorem \ref{thm-Brenier} that $c_{xyy}>0$. Then, the optimal upper bound is given by
 \be
 P_2(\mu,\nu)
 &=&
 \int_0^ \infty \Big[q(x)\ln^2\Big({T_u(x) \over x}\Big)
                     +(1-q)(x)\ln^2 \Big({T_d(x) \over x}\Big)
                \Big]\mu(dx),
 \ee
where $q$ is set to an arbitrary value on $D$. In Figure \ref{SuperVS}, we have plotted $\varphi_*,\psi_*$ and $h_*$ with marginal distributions $\mu_0=e^{\Nc(-\sigma_1^2/2,\sigma_1^2)}$ and $\nu_0=e^{\Nc(-\sigma_2^2/2,\sigma_2^2)}$, $\sigma_1^2=.04<\sigma_2^2=.32$. We recall that the corresponding maps $T_d,T_u$ are plotted in Figure \ref{Ex1}. The expression for $\psi_*$ is
 \b*
 \psi_*'(x)
 =
 {2 \over x} \ln \left({x \over T_u^{-1}(x)}\right)
 + 2 \int_{x_0}^{T_u^{-1}(x)} {
 \ln \left({T_u(\xi) \over T_d(\xi)}\right)
     \over
     \xi(T_u(\xi)-T_d(\xi))}
 d\xi.
 \e*
In particular, $\psi_*''(x)={2 \over x^2}$ for all $x\leq m_1$.

\begin{figure}[tbp]
\begin{center}
\includegraphics[width=7cm,height=7cm]{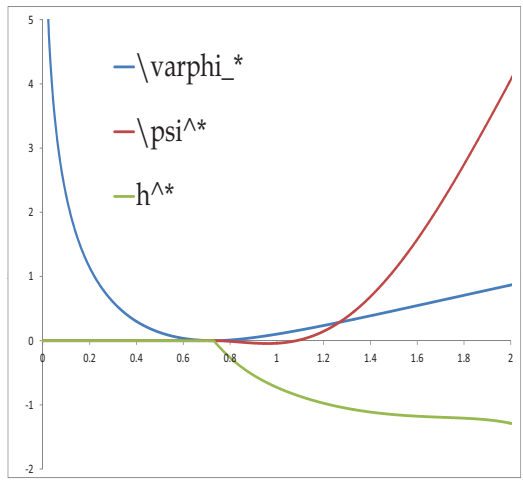}
\end{center}
\caption{Superreplication strategy for a $2$-period variance swap
given  two log-normal densities with variances $0.04$ and $0.32$.}
\label{SuperVS}
\end{figure}
\end{Example}

\begin{Example}[$c(x,y)=-\left({y \over x}\right)^p$, $p>1$, and $\mu,\nu$ have support in $(0,\infty)$] This payoff function also
satisfies the condition of Theorem \ref{thm-Brenier} that $c_{xyy}>0$. The upper bound is
 \b*
 P_2(\mu,\nu)
 =
 -\int_0^\infty \Big[q(x)\Big(\frac{T_u(x)}{x}\Big)^p
                     +(1-q)(x)\Big(\frac{T_d(x)}{x}\Big)^p
                \Big]
                \mu(dx).
 \e*

 \end{Example}

\section{The $n-$Marginals Martingale Transport}
\label{sect:multimarginals}

In this section, we provide a direct extension of our results to the martingale transportation problem under
finitely many marginals constraint. Fix an integer $n\ge 2$, and let $X=(X_1,\ldots,X_n)$ be a vector of $n$
random variables denoting the prices of some financial asset at dates $t_1<\ldots<t_n$. Consider the probability
measures
$\mu=(\mu_1,\ldots,\mu_n)\in(\Pc_{\R})^n$ with
$\mu_1\preceq\ldots\preceq\mu_n$ in the convex order and
 \b*
 \int|\xi|\mu_i(d\xi)<\infty
 &\mbox{and}&
 \int \xi\mu_i(d\xi)=X_0,
 ~~\mbox{for all}~~
 i=1,\ldots,n.
 \e*
Similar to the two-marginals case, we introduce the set
 \b*
 \Mc_n(\mu)
 &:=&
 \big\{\P\in\Pc_n(\mu): X~\mbox{is a $\P-$martingale}
 \big\},
 \e*
where $\Pc_n(\mu)$ was defined in (\ref{P0n}). In the present martingale version, we introduce the one-step ahead
martingale transport maps defined by means of the $n$ pairs of maps $(T_d^i,T_u^i)$:
 \begin{eqnarray}
 T_*^i(x_i,.)
 &:=&
 \1_{D_i}\delta_{\{x_i\}}
 +\1_{D_i^c}
   \big(q_i(x_i)\delta_{T_u^i(x_i)}
        +(1-q_i)(x_i)\delta_{T_d^i(x_i)}
   \big),
 \end{eqnarray}
where $q_i(\xi):=(\xi-T_d^i(\xi))/(T_u^i-T_d^i)(\xi)$ for
$\xi\in D_i^c$, and $(D_i,T^i_d,T^i_u)_{i=1,\ldots, n-1}$ are defined as in Subsection \ref{sect-explicitconstruction}
with the pair $(\mu_i,\mu_{i+1})$.

The $n-$marginals martingale transport problem is defined by:
 \b*
 P_n(\mu)
 =
 \sup_{\P\in\Mc_n(\mu)}
 \E^{\P}[c(X)],
 \e*
where the map  $c:\R^n\longrightarrow\R$ is of the form
 \b*
 &c(x_1,\ldots,x_n)
 =
 \sum_{i=1}^{n-1} c^i(x_i,x_{i+1})
 \e*
for some upper semicontinuous functions
$c^i:\R\times\R\longrightarrow\R$ with linear growth (or the
condition (\ref{othercond})), $i=1,\ldots,n-1$.

The dual problem is defined by
 \b*
 D_n(\mu)
 &:=&
 \inf_{(u,h)\in\Dc_n}\sum_{i=1}^n\mu_i(u_i),
 \e*
where $u=(u_1,\ldots,u_n)$ with components $u^i:\R\longrightarrow
\R$, and $h=(h_1,\ldots,h_{n-1})$ with components
$h_i:\R^i\longrightarrow\R$, taken from the set of dual variables:
 \b*
 &\Dc_n
 :=
 \big\{(u,h):(u_i)^+\in\L^1(\mu_i),
              h_i\in\L^0(\R^i),
              ~\mbox{and}~
              \oplus_{i=1}^n u_i
              + \sum_{i=1}^{n-1} h_i^{\otimes^i}
              \ge c
 \big\}.
 \e*
Here, $\oplus_{i=1}^n u_i(x)=\sum_{i\le n}u_i(x_i)$ and $h_i^{\otimes^i}(x)=h_i(x_1,\ldots,x_i)(x_{i+1}-x_i)$.

Similar to the two-marginals problems, the weak duality inequality $P_n(\mu)\le D_n(\mu)$ is obvious, and we shall
obtain equality in the following result under convenient conditions.

To derive the structure of the optimal hedging strategy, we shall consider the two-marginals $(\mu_i,\mu_{i+1})$
problems with coupling functions $c^i$. By Theorem \ref{thm-Brenier}, we have for $i=1,\ldots,n-1$:
 $$
 P^i_2(\mu_i,\mu_{i+1})
 :=
 \sup_{\P\in\Mc(\mu_i,\mu_{i+1})}\E^\P[c^i(X,Y)]
 =
 \inf_{(\varphi,\psi,h)\in\Dc^i_2}
 \{\mu_i(\varphi)+\mu_{i+1}(\psi)\}
 =\mu_{i}(\varphi_i^*)+\mu_{i+1}(\psi_i^*),
 $$
where $\Dc^i_2$ is defined as in \eqref{defD} with $c^i$ substituted
to $c$, and $(\varphi_i^*,\psi_i^*,h_i^*)\in\Dc^i_2$ are defined as
in \eqref{defh}-\eqref{defg}-\eqref{deff} with $c^i$ substituted to
$c$ and $(T_u^i,T_d^i)$ substituted to $(T_u,T_d)$. Finally, we
define:
 \b*
 u^*_i(x_i)
 :=
 \1_{\{i<n\}}\varphi^*_i(x_i)
 +\1_{\{i>1\}}\psi^*_{i-1}(x_i),
 &i=1,\ldots,n,&
 \e*
and $u^*:=\big(u^*_1,\ldots,u^*_n\big)$, $h^*:=\big(h^*_1,\ldots,h^*_{n-1}\big)$.

\begin{Theorem} \label{thm-Brenier-n}
Suppose $\mu_1\preceq\ldots\preceq\mu_n$ in convex order, with
finite first moment, $\mu_1,\ldots,\mu_{n-1}$ have no atoms, and let Assumption \ref{assum-rightaccumulation} hold true for $\delta F=F_{\mu_{i+1}}-F_{\mu_i}$, for all $1\le i<n$. Assume further that

$\bullet$ $c^i$ have linear growth, that the cross derivatives $c^i_{xyy}$ exist and satisfy $c^i_{xyy}>0$,

$\bullet$ $\varphi_i^*,\psi_i^*$ satisfy the integrability
conditions $(\varphi_i^*)^+ \in\L^1(\mu_i)$,
$(\psi_i^*)^+\in\L^1(\mu_{i+1})$.

\no Then, the strong duality holds, the transference map $\P^*_n(dx)=\mu_1(dx_1)\prod_{i=1}^{n-1}T_*^i(x_i,dx_{i+1})$
is optimal for the martingale transportation problem $P_n(\mu)$, and $(u^*,h^*)$ is optimal for the dual problem
$D_n(\mu)$, i.e.
 \b*
 \P^*_n\in\Mc_n(\mu),
 ~~
 (u^*,h^*)\in\Dc_n,
 ~~\mbox{and}
 &\E^{\P^*_n}[c(X)]=P_n(\mu)=D_n(\mu)=\sum_{i=1}^n\mu_i(u^*_i).&
 \e*
\end{Theorem}

\proof Clearly, we have $\P^*_n\in\Mc_n(\mu)$, which provides the
inequality $\E^{\P^*_n}[c(X)]\le P_n(\mu)$. We next observe that
$(u^*,h^*)\in\Dc_n$ from our construction. Then $D_n(\mu)\le
\sum_{i\le n}\mu_i(u_i^*)= \E^{\P^*_n}[c(X)]$. The required result
follows from the weak duality inequality $P_n(\mu)\le D_n(\mu)$. \ep

\begin{Remark}\label{rem:lowerboundn}
The optimal lower bound for a coupling function as in Theorem \ref{thm-Brenier-n} is attained
by the mirror solution introduced in Remark \ref{rem-mirror}.
\end{Remark}

\begin{Example}[Discrete monitoring variance swaps] This is a continuation of our Example \ref{ex-Varswap}. Suppose that
$\mu_1\preceq\ldots\preceq\mu_n$ have support in $(0,\infty)$ with mean
$X_0$, and let $c(x_1,\ldots,x_n):=\sum_{i=1}^n \big(\ln {x_i \over
x_{i-1}}\big)^2$. Then:
 \b*
 P_n(\mu)
 =
 \int\Big(\ln {\xi \over X_0}\Big)^2\mu_1(d\xi)
 +
 \sum_{i=1}^{n-1}
 \int_0^\infty \Big[q_i(\xi)
                    \Big(\ln\frac{T^i_u(\xi)}{\xi}\Big)^2
                    +(1-q_i)(\xi)
                    \Big(\ln\frac{T^i_d(\xi)}{\xi}\Big)^2
               \Big]\mu_i(d\xi).
 \e*
This optimal bound depends on all the marginals. The optimal lower bound is attained by our mirror solution, see Remark
\ref{rem:lowerboundn}.
\end{Example}

\begin{Remark}
 In particular, their argument holds whenever
 $c(x,y)$ which satisfies $c(x,x)=0=c_y(x,x)$, $(x-y)c_{xy}+c_x>0$ and our generalized Spence-Mirrlees condition $c_{xyy}>0$. Note that apart from the last condition, these requirements on $c$ are not preserved by the transformation in Remark \ref{symmetrytransformation}.
\end{Remark}

\begin{Remark}
In a related robust hedging problem, Hobson and Klimmek \cite{HobsonKlimmek}, derived an optimal upper bound for a derivative $c(x_1,\ldots,x_n)=\sum_{i=1}^{n-1}c^0(x_i,x_{i+1})$. The difference with our problem above is that they are only given the marginal distribution $\mu_n$ for $X_n$. See also Kahale \cite{kah}. We would like to emphasize that \cite{HobsonKlimmek} assume the variance Kernel $c_0$ to satisfy the conditions $c^0(x,x)=c^0_y(x,x)=0$, $(x-y)c_{xy}+c_x>0$, together with our Spence-Mirrlees condition $c_{xyy}>0$. In the context of our problem with finitely many given marginals $\mu_1,\ldots,\mu_n$, notice that, apart from the Spence-Mirrlees condition, none of these requirements are preserved by the transformation of Remark \ref{symmetrytransformation}.
\end{Remark}

\section{Proof of the main results}
\label{sect:proofs}

\subsection{Construction of the left-monotone map}
\label{sect-construction}

This section is devoted to the proof of Theorem \ref{thm-defT}.
We first motivate the definition of the maps $T_d$ and $T_u$ through the functions $g$ and $G$. In this heuristic discussion, we consider the simple case of one single maximizer $m_1$ with $\delta F$ strictly increasing before $m_1$, and we ignore the possible jumps of $F_\nu$.

The first observation about our construction is that for a point $y\in\R$, there are two alternatives:

$\bullet$ either $y\in (-\infty,m_1]$; then $\P_*[Y\in dy]
 =
 dF_\mu(y)+\E\big[(1-q)(X)\1_{\{T_d(X)\in dy\}}\big]$, and the requirement that $Y\sim_{\P_*}\nu$ together with
  the decrease of $T_d$ imply that
 \b*
 d(\delta F\circ T_d)
 &=&
 -(1-q)dF_\mu;
 \e*
in particular, in order for $T_d$ to be well-defined, it has to be valued in the domain of increase of $\delta F$,

$\bullet$ or $y\in (m_1,\infty)$, then $
 \P_*[Y\in dy]
 =
 \E\big[q(X)\1_{\{T_u(X)\in dy\}}\big]$, and the requirement that $Y\sim_{\P_*}\mu_2$ together with the increase
  of $T_u$ imply that
 \b*
 d(F_\nu\circ T_u)
 &=&
 qdF_\mu.
 \e*
Direct manipulation of these two equations implies that
$ d\delta F(T_d)=-dF_\mu+dF_\nu(T_u)$.
Since $T_d(m_1)=T_u(m_1)=m_1$, this implies that:
 \b*
 F_\nu\big(T_u(x)\big)
 &=&
 F_\mu(x)+\delta F\big(T_d(x)\big),
 \e*
i.e. $T_u=g(.,T_d)$ as in (\ref{TdTu1}), (\ref{TdTu2}),
and (\ref{TdTui}).

Also, as a consequence of this relation, we see that the requirement $T_u(x)\ge x$ implies that $\delta F(x)\le\delta F(T_d(x))$. Consequently, the choice of the break point $m_1$ as the maximizer of $\delta F$ is necessary.

We next substitute $q$ and $T_u$ in the martingale condition:
 \b*
 xdF_\mu
 &=&
 T_u qdF_\mu+T_d(dF_\mu-qdF_\mu)
 \;=\;
 g(.,T_d)d[F_\mu+\delta F(T_d)]-T_dd\delta F(T_d).
 \e*
This implies that, for $x>m_1$:
 \b*
 [x-F_\nu^{-1}\circ F_\mu(x)]dF_\mu
 =
 d\Big\{\int_0^{\delta F(T_d)}F_\nu^{-1}(F_\mu(x)+y)dy
  \Big\}
 -T_dd\delta F(T_d).
 \e*
Integrating from $m_1$ to $x$, and using the condition $T_d(m_1)=m_1$, this provides:
 \b*
 G(T_d(x),x)-G(m_1,m_1)=0
 \e*
where
 \b*
 G(t,x)
 &:=&
 \int_{-\infty}^x[F_\nu^{-1}\circ F_\mu(\xi)-\xi]dF_\mu(\xi)
         +\int_0^{\delta F(t)}F_\nu^{-1}(F_\mu(x)+y)dy
         -\int_{(-\infty,t]}\xi d\delta F(\xi)
 \\
 &=&
 \int_{(-\infty,F_\nu^{-1}\circ F_\mu(x)]}\xi dF_\nu(\xi)
 -\int_{-\infty}^x\xi dF_\mu(\xi)
 +\int_{(-\infty,t]}[g(x,\xi)-\xi] d\delta F(\xi),
 \e*
in agreement with our definition of $G^A$ in \eqref{GA} for $A=(-\infty,m_1]$. We finally notice by direct computation that $G(m_1,m_1)=0$, so that $T_d$ must satisfy the equation $G(T_d(x),x)=0$ for all
$x \ge m_1$.

\vspace{5mm}

\no {\bf Proof of Lemma \ref{lem-stepi}} (i) Since $\delta F$ is
strictly increasing on $A$, we see that $F_\nu$ is strictly
increasing in $A$. Therefore, for $t<m\le x$, $t\in A$ we have
$g(x,t)-t>g(t,t)-t=0$, implying that $t\longmapsto G^A(t,x)$ is strictly increasing in $t$ on the set $A$.

We next verify that $G^A(m,x)>0$ as long as $\delta F(m)>\delta F(x)$. Denoting by $d_x$ the differential with respect to the $x-$variable, we compute by using the conditions on the set $A$ that
 \b*
 d_xG^A(m,x)
 &=&
 \Big(F_\nu^{-1}\circ F_\mu(x)-x
               +\int_{(-\infty,m]}\partial_xg(x,\xi)\1_A(\xi)d\delta F(\xi)
 \Big)dF_\mu(x)
 \\
 &=&
 \Big(g(x,m)-x+\int_{\bar A_m}dg(x,\xi)\Big)dF_\mu(x)
 \;\ge\;
 (g(x,m)-x)dF_\mu(x),
 \e*
since $g(x,\xi)=\phi(\delta F(\xi))$ where, for fixed $x$, the function $y\mapsto\phi(y):=F_\nu^{-1}(\delta F(x)+y)$ is nondecreasing. Since $F_\mu$ strictly increases at the right of $m$, and $G^A(m,m)=0$, this shows that $G^A(m,x)>0$ as long as $g(x,m)-x>0$ or, equivalently, $\delta F(m)>\delta F(x)$.

Then, in order to establish the existence and uniqueness of
$t^A(x,m)$, it remains to verify that
 \b*
 \gamma(x):=G^A(-\infty,x)
 =\int_{(-\infty,F_\nu^{-1}\circ F_\mu(x)]} \xi dF_\nu(\xi)
  -\int_{-\infty}^x\xi dF_\mu(\xi)
 <0
 &\mbox{for}&
 \delta F(x)\le \delta F(m).
 \e*
Let $\bar x_0:=\inf\{x:\delta F(x)>0\}$. Clearly, $\bar x_0<m\le x$, and $\gamma=0$ on $(-\infty,\bar x_0)$, $\gamma(r_\nu)=0$. Moreover, $\gamma$ is flat on Supp$(F_\mu)^c$, where Supp$(F_\mu)$ is a support of $F_\mu$, and we see by direct differentiation that $\gamma$ is absolutely continuous with respect to $\mu$ with:
 \b*
 d\gamma(x)
 &=&
 (F_\nu^{-1}\circ F_\mu(x)-x)dF_\mu(x),
 \e*
implying that $d\gamma<0$ at the right of $\bar x_0$, by the (strict) convex-order property ($\mu\preceq\nu$) implied by the strict increase of $\delta F$ on $A$. Furthermore, let $x^*$ be any possible local maximizer of $\gamma$. By the fact that $\gamma$ is flat off $\mbox{Supp}(F_\mu)$, we may assume that $x^*$ is either an interior point of $\mbox{Supp}(F_\mu)$ or $x^*$ is a left accumulation point of $\mbox{Supp}(F_\mu)$. In both cases, it follows from the first order condition that
 \b*
 F_\nu^{-1}\big(F_\mu(x^*)-\big)
 \le
 x^*
 \le
 F_\nu^{-1}\big(F_\mu(x^*)\big).
 \e*
If $F_\nu^{-1}$ is continuous at the point $F_\mu(x^*)$, then $\delta F(x^*)=0$, and it follows from the definition of $\gamma$ that
 \b*
 \gamma(x^*)
 &=&
 \int_{(-\infty,x^*]}\xi d\delta F(\xi)
 \;=\;
 -\int_{(-\infty,x^*]}(x^*-\xi)d\delta F(\xi)
 .
 \e*
By the (strict) convex-order property, this implies that $\gamma(x^*)<0$.

In the alternative case that $F_\nu^{-1}$ jumps at the point $F_\mu(x^*)$, notice that $F_\nu$ is flat at the right of $F_\nu^{-1}\circ F_\mu(x^*)$, and therefore the conclusion $\gamma(x^*)<0$ holds true in this case as well. Consequently, $\gamma<0$ on $(\bar x_0,r_\mu)$. Since $x\ge m>\bar x_0$,
this provides the required strict inequality.

\no (ii) Suppose $\bar x(m)<\infty$. Since $\delta F$ is strictly
increasing on $A$, the inequalities \eqref{ineq-tA} follow from the
definition of $\bar x(m)$.

It remains to prove that $\delta F$ strictly increases in a right
neighborhood of $\overline{x}(m)$ whenever $\overline{x}(m)<\infty$.
By definition, we have $t^A(x,m)>(\delta F)^{-1}\circ\delta F(x)$ on
$(m,\overline{x}(m))$, and $t^A(\overline{x}(m),m)\le(\delta F)^{-1}\circ\delta F(\overline{x}(m))$, where
$\delta F^{-1}$ denotes the inverse function of
$\int^._{-\infty}\1_Ad\delta F$. We denote $h(x,m):=G^A((\delta
F)^{-1}\circ\delta F(x), x)$, and we compute that
$d_xh(x,m)=[x-(\delta F)^{-1}\circ\delta F(x)]d\delta
F(x)$. Since $x>(\delta
F)^{-1}\circ\delta F(x)$ whenever $x>m$, we see that $h(.,m)$
decreases down from zero on the right neighborhood of $x=m$
(confirming that $t^A(x,m)>(\delta F)^{-1}\circ\delta F(x)$ near
$m$), and has the same maximum and minimum points as the function
$\delta F$. Since $h$ must be increasing at a right neighborhood of
$\overline{x}(m)$, it follows that $\delta F$ has the same property.
\ep

\vspace{5mm}

\no {\bf Proof of Theorem \ref{thm-defT}} (i) By construction,
the probability measure $\P_*$ satisfies the
left-mo\-no\-to\-ni\-ci\-ty property of Definition
\ref{def-leftmonotone}. In the rest of this proof, we verify that
$\P_*\in\Mc_2(\mu,\nu)$. In particular, by the uniqueness result of
Beiglb\"{o}ck and Juillet \cite{BeiglbockJuillet} (Theorem 1.5 and
Corollary 1.6), this would imply that $\P_*$ is the unique left
monotone transport plan.

First, by the definition of $\P_*$ in (\ref{T*}),
$X\sim_{\P_*}\mu$, and $\E^{\P_*}[Y|X]=X$. It remains to verify that $Y\sim_{\P_*}\nu$. We argue as in the
beginning of Section \ref{sect-construction} considering separately the following alternatives for any point $y\in\R$:
\begin{itemize}
\item[Case 1:] $y=y_d\in D\cap B_0$ corresponds to some point $x$ such that $y_d=T_d(x)$, and we see from
the definition of $\P_*$ that:
  \b*
  \P_*[Y\in dy_d]
  =
  dF_\mu\big(T_d(x)\big)-(1-q)dF_\mu(x)
  &\mbox{and}&
  dF_\nu\big(T_u(x)\big)=qdF_\mu.
  \e*
Then, $\P_*[Y\in dy_d]=dF_\mu(y_d)-dF_\mu(x)+dF_\nu(T_u(x))$.
Since $T_u(x)=g(x,T_d(x))$, this provides $\P_*[Y\in
dy_d]=F_\nu(dy)$ by direct substitution.
\item[Case 2:] $y=y_u\in D^c$ corresponds to some $x$ such that $y_u=T_u(x)$, and we see from the
 definition of $\P_*$ that:
 $$
 \P_*[Y\in dy_u]
 =
 qdF_\mu(x)
 =
 d\delta F(T_d(x))+dF_\mu.
 $$
Using again the expression of $T_u$ in terms of $T_d$, it follows that
 $$
 \P_*[Y\in dy_u]
 =
 d\big(F_\nu\circ T_u(x)-F_\mu(x)\big)+dF_\mu(x)
 =
 dF_\nu(x).
 $$
\item[Case 3:] At a point of discontinuity of $T_u$ or $T_d$, the above cases 1 and 2 are immediately adapted
to account for the point mass.
\item[Case 4:] In the remaining alternative $y\in D\setminus B_0$, we observe that the function $\delta F$ is
flat near $y$, and there is no $x\neq y$ such that $T_d(x)=y$ or $T_u(x)=y$. Then, it follows from the
definition of $\P_*$ that:
 $$\P_*[Y\in dy]=dF_\mu(y)=dF_\nu(y).$$
\end{itemize}
\no (ii) Differentiating the integral equation defined by $G^A$ at a continuity point of $T_d$, we see that:
 \b*
 0
 &=&
 -\big[F_\nu^{-1}\circ F_\mu(x)-x\big]dF_\mu(x)
 +\big[g(x,T_d(x))-F_\nu^{-1}\circ F_\mu(x)\big]dF_\mu(x)
 \\
 &&
 +\big[g(x,T_d(x))-T_d(x)\big]d\delta F(T_d(x))
 \\
 &=&
 \big[g(x,T_d(x))-x\big]dF_\mu(x)
 +\big[g(x,T_d(x))-T_d(x)\big]d\delta F(T_d(x)).
 \e*
Since $T_u=g(.,T_d)$ this is the required ODE. The ODE for $T_u$ is
obtained by using the relation $T_u=g(.,T_d)$.
\ep

\subsection{The optimal semi-static hedging strategy}

We start by following the same line of argument as in the proof of
Theorem \ref{thm-Frechet} in order to identify the semi-static
hedging strategy introduced in (\ref{defh}-\ref{defg}-\ref{deff}).
Our objective is then to construct a pair
 \be\label{goal}
 (\varphi_*,\psi_*,h_*)\in\Dc_2
 &\mbox{such that}&
 \mu(\varphi_*)+\nu(\psi_*)=
 \E^{\P_*}[c(X,Y)].
 \ee
This will provide equality in (\ref{weakduality}) with the
optimality of $\P_*$ for the optimal transportation problem $P_2$
and the optimality of $(\varphi_*,\psi_*,h_*)$ for the dual problem
$D_2$.

By the definition of the dual set $\Dc_2$, we observe that the
requirement (\ref{goal}) is equivalent to
 \be\label{perfectreplication}
 \varphi_*(X)+\psi_*(Y)+h_*(X)(Y-X)-c(X,Y)=0,
 &\P_*-\mbox{a.s. for some function}&
 h_*,
 \ee
and that the function $\varphi_*$ is determined by:
 \begin{equation}\label{f=max}
 \varphi_*(x)
 =
 \max_{y\in\R}H(x,y),
 ~~\mbox{where}~~
 H(x,y):=c(x,y)-\psi_*(y)-h_*(x)(y-x),
 ~~x,y\in\R.
 \end{equation}
The perfect replication property (\ref{perfectreplication}), is
equivalent to:
 \be\label{f}
 \varphi_*(x)
 &=&
 q(x)(c(x,.)-\psi_*)\circ T_u(x)
 +(1-q(x))(c(x,.)-\psi_*)\circ T_d(x),
 \\
 h_*(x)
 &=&
 \frac{(c(x,.)-\psi_*)\circ T_u(x)
        -(c(x,.)-\psi_*)\circ T_d(x)}
      {(T_u-T_d)(x)}
 ~~\mbox{for}~~x\in D^c,
 \label{h}
 \ee
where we observe that we may choose $h_*$ arbitrarily on $D$.

It remains to determine $\psi_*$ by using the static superhedging
condition (\ref{f=max}). Since $T_u$ and $T_d$ are maximizers in
(\ref{f=max}), it follows from the first-order condition that
 \be
 &\psi_*'\circ T_u(x)
 =
 c_y(x,T_u(x))-h_*(x),
 ~~
 \psi_*'\circ T_d(x)
 =
 c_y(x,T_d(x))-h_*(x),
 ~~x\in D^c,
 &\label{gx*+}
 \\
 &\mbox{and}~~
 \psi_*'(x)
 =
 c_y(x,x)-h_*(x)
 ~~\mbox{for}~~x\in D.
 &\label{gx*-}
 \ee
We now determine $h_*$. Differentiating (\ref{h}), and using
(\ref{gx*+}), we see that for $x\in D^c$:
 \b*
 h_*'(x)
 &=&
 \frac{d}{dx}\Big\{\frac{c(x,T_u)-c(x,T_d)}{T_u-T_d}
             \Big\}
 +\frac{T_u'-T_d'}{T_u-T_d}\;
  \frac{\psi_*(T_u)-\psi_*(T_d)}{T_u-T_d}
 \\
 &&
 +\frac{T_d'\big[c_y(x,T_d)-h_*]-T_u'\big[c_y(x,T_u)-h_*]}
       {T_u-T_d}
 \e*
Then, direct calculation leads to the expression of $h_*'$ on $D^c$
reported in (\ref{defh}). Since $T_d$ and $T_u$ take values in $D$
and $D^c$, respectively, and $h_*$ is determined by the last two
equations, we see that equation (\ref{gx*+}) determines $\psi_*$ on
$\R$. We finally observe that by (\ref{gx*+}) and (\ref{gx*-}), we
have for $x\in D$ that $\psi_*'(x)=c_y(T_d^{-1}(x),x)-h_*\circ
T_d^{-1}(x)=c_y(x,x)-h_*(x)$, which completes the definition of
$h_*$, up to an irrelevant constant, on $D$.

\subsection{Proof of Theorem \ref{thm-Brenier}}

Following the line of argument of the proof of Theorem
\ref{thm-Frechet}, we see from the weak duality (\ref{weakduality})
that
 \b*
 \E^{\P_*}[c(X,Y)]
 \le P_2(\mu,\nu)
 \le D_2(\mu,\nu).
 \e*
Then, the proof of Theorem \ref{thm-Brenier} is completed by the
following result.

\begin{Lemma}
Let $\mu,\nu$ be as in Assumptions \ref{assum-munu} and \ref{assum-rightaccumulation}, and suppose that
the payoff function $c$ satisfies $c_{xyy}>0$. Then
$\varphi_*\oplus\psi_*+h_*^\otimes\ge c$.
\end{Lemma}

\no {\bf Proof} (i) We first verify that $T_u$ and $T_d$ satisfy the second order condition for a local maximum on $D^c$. Differentiating (\ref{gx*+}), and using the expression of $h_*'$ in (\ref{defg}), it follows from the condition $c_{xyy}>0$ that, in the distribution sense,
 \b*
 H_{yy}(.,T_u)T_u'
 \;=\;
 \big[c_{yy}(.,T_u)-\psi_*''\circ T_u\big]T_u'
 &=&
 \frac{c_x(.,T_u)-c_x(.,T_d)}
      {T_u-T_d}
 -c_{xy}(.,T_u)
 \;<\;
 0
 \\
 H_{yy}(.,T_d)T_d'
 \;=\;
 \big[c_{yy}(.,T_d)-\psi_*''\circ T_d\big]T_d'
 &=&
 \frac{c_x(.,T_u)-c_x(.,T_d)}
      {T_u-T_d}
 -c_{xy}(.,T_d)
 \;>\;
 0,
 \e*
on $D^c$. By the nondecrease of $T_u$ and the nonincrease of $T_d$, this implies that $H_{yy}(.,T_u)<0$ and $H_{yy}(.,T_d)<0$.
\\
(ii) We next show that $y\longmapsto H(.,y)$ is increasing before $T_d$, and decreasing after $T_u$. In particular,
this implies that:
 \b*
 \varphi_*(x)
 =\max_{y\in[T_d(x),T_u(x)]}H(x,y)
 &\mbox{for all}&
 x\in\R.
 \e*
Set $y:=T_u(x)$, let $m_i$ be the local maximum from which $(T_d,T_u)(x)$ is constructed, and consider an arbitrary $y'=T_u(x')>y$ for some $x'>x$. We only report the proof for the case $x'\in (m_j,x_j]$ for some $j\ge i$, the remaining cases are treated similarly. Recalling that $H_y(x,T_u(x))=0$, we decompose
 \b*
 H_y(x,y')
 &=&
 H_y(x,y')-H_y(x,m_j)
 +\sum_{i+1}^j (A_k+B_k),
 \e*
where
 $$
 A_k:=H_y(x,m_k)-H_y(x,x_{k-1}),
 ~~
 B_k:=H_y(x,x_{k-1})-H_y(x,m_{k-1}\wedge T_u(x)).
 $$
We next compute from the expression of $h_*$ in (\ref{defh}) that:
 \b*
 H_y(x,y')-H_y(x,m_j)
 &=&
 \int_{m_j}^{y'}\big[c_{yy}(x,\xi')-\psi''(\xi')\big]d\xi'
 \\
 &\le&
 \int_{m_j}^{y'}\big[c_{yy}(x,\xi')-c_{yy}(T_u^{-1}(\xi'),\xi')\big]d\xi'
 \\
 &=&
 \int_{m_j}^{y'}
 \int_x^{T_u^{-1}(\xi')}
 c_{xyy}(\xi,\xi')d\xi d\xi'
 <0,
 \e*
where the second inequality follows from the second order condition verified in (i). Similarly, we compute that
 \b*
 A_k
 &=&
 \int_{x_{k-1}}^{m_k}
 \big[c_{yy}(x,\xi')-\psi''(\xi')\big]d\xi'
 \\
 &\le&
 \int_{x_{k-1}}^{m_k}
 \big[c_{yy}(x,\xi')-c_{yy}(T_d^{-1}(\xi'),\xi')\big]d\xi'
 \\
 &=&
 -\int_{x_{k-1}}^{m_k}
 \int_x^{T_d^{-1}(\xi')}c_{xyy}(\xi,\xi')d\xi d\xi'
 < 0,
 \e*
where we used again the second order condition verified in (i). Finally,
 \b*
 B_k
 &=&
 \int_{m_{k-1}\vee T_u(x)}^y
 \big[c_{yy}(x,\xi')-\psi_*''(\xi')\big]d\xi'
 \\
 &\le&
 \int_{m_{k-1}\vee T_u(x)}^y
 \big[c_{yy}(x,\xi')-c_{yy}(T_u^{-1}(\xi'),\xi')\big]d\xi'
 \\
 &=&
 -\int_{m_{k-1}\vee T_u(x)}^y
 \int_x^{T_u^{-1}(y')}c_{xyy}(\xi,\xi')d\xi d\xi'
 < 0.
 \e*
A similar argument also shows that $H_y(x,y')<0$ for $y'<T_d(x)$.
\\
(iii) We next show that $H(.,T_d)=H(.,T_u)$. Denote $\delta H:=H(.,T_u)-H(.,T_d)$, and compute:
 \b*
 \delta H'
 &:=&
 c_x(.,T_u)-c_x(.,T_d)-(T_u-T_d)h_*'
 \\
 &&
 +\big[c_y(.,T_u)-\psi_*'(T_u)-h_*\big]T_u'
 -\big[c_y(.,T_d)-\psi_*'(T_d)-h_*\big]T_d'
 \e*
in the distribution sense. By definition of $\psi_*$ and $h_*$, it follows that $\delta H'=0$ at any continuity point. Since $\delta H$ is continuous by our construction, see \eqref{deltaHcontinuous}, this shows that $\delta H(x)=\delta H(m_i)=0$, where $m_i$ is the local maximizer from which $(T_d,T_u)(x)$ is defined.
\\
(iv) We finally show that $T_u$ and $T_d$ are global maximizers of $y\longmapsto H(.,y)$. Let $x\in D^c$,
and denote by $m$ the local maximizer from which $T_d(x)$ and $T_u(x)$ are constructed. For fixed $T=T_u(t)\in\big(m,T_u(x)\big)$, it follows from similar calculations as in the previous step that
 \b*
 \partial_x
 \big\{H(.,T_u)-H(.,T)\big\}
 &=&
 c_x(.,T_u)-c_x(.,T)-(T-T_d)h_*'
 \\
 &=&
 (T_u-T)\Big(\frac{c_x(.,T_u)-c_x(.,T)}
                  {T_u-T}
             -\frac{c_x(.,T_u)-c_x(.,T_d)}
                  {T_u-T_d}
        \Big)
 >0
 \e*
by the condition $c_{xyy}>0$. Then $H(.,T_u)-H(.,T)=\int_t^.\partial_x
 \big\{H(.,T_u)-H(.,T)\big\}>0$.

By a similar calculation, we also show that
$H\big(x,T_d(x)\big)-H(x,T)\ge 0$ for all $T\in(T_d(x),m)$, thus
completing the proof that $T_d$ and $T_u$ are global maximizers of $y\longmapsto H(.,y)$.
\ep

\section{Complement: Some examples of martingale measures given marginals}
\label{sect-expMc}

\subsection{Local volatility model}

A first example of a martingale measure fitted to two marginal
distributions $\mu_{t_1}$ and $\mu_{t_2}$, corresponding to the
maturities $t_1<t_2$, is given by the Dupire local volatility model
(in short LV) \cite{dup}. We first define an interpolation
$(\mu_t)_{t\in[t_1,t_2]}$ which does not violate the no-arbitrage
condition, i.e. which obeys to the convex ordering condition. This
can be achieved by introducing the implied Black-Scholes accumulated
variances $\varpi(t_i,K)$, defined by
BS$\big(K,\varpi(t_i,K)\big)={\cal C}(t_i,K):=\int
(\xi-K)^+\mu_{t_i}(d\xi)$, where BS denotes the Black-Scholes
formula
 \b*
 {\rm BS}(K,v)
 &:=&
 X_0 \mathbf{N}\Big(\frac{\ln{(X_0/K)}}{\sqrt{v}}+\frac{\sqrt{v}}{2}
               \Big)
 - K\; \mathbf{N}\Big(\frac{\ln{(X_0/K)}}{\sqrt{v}}-\frac{\sqrt{v}}{2}
               \Big),
 \e*
with $\mathbf{N}$ the c.d.f. of the standard Normal distribution, and for $t\in[t_1,t_2]$:
 $$
 \varpi(t,K)
 =
 {t_2-t \over t_2-t_1}\varpi(t_1,K)
 +{t-t_1 \over t_2-t_1}\varpi(t_2,K),
 ~
 {\cal C}(t,K)
 \!:=\!
 {\rm BS}\big(K,\varpi(t,K)\big)
 \!=\!\int (\xi-K)^+\mu_{t}(d\xi).
 $$
The Dupire LV model corresponding to this interpolation is defined by the SDE:
 \b*
 dX_t
 =
 X_t\sigma_\mathrm{loc}(t,X_t) dW_t
 &\mbox{with}&
 \sigma_\mathrm{loc}(t,K)^2
 :=
 2\;{\partial_t {\cal C}(t,K)
 \over
 \partial_K^2 {\cal C}(t,K)},
 \e*
whenever $\sigma_\mathrm{loc}$ is well-defined and induces a well-defined weak solution for the above SDE. In general, $\sigma_\mathrm{loc}$ is a measure with poor regularity. A rigorous adaptation of this solution, by convenient regularization of $\sigma_{loc}$, is provided by
Hirsch and Roynette \cite{hir0}, resulting in a new proof of the Kellerer theorem \cite{kel}.

A natural extension of such a LV model is given by the so-called
local stochastic volatility model in which $X_t$ satisfies a
non-linear McKean SDE (\cite{guy}):
 \b*
 dX_t
 =
 X_t\;
 {\sigma_\mathrm{loc}(t,X_t) \over \sqrt{\E[a_t^2|X_t]}}a_tdW_t
 \e*
with $a_t$ a (possibly multi-dimensional) It\^o diffusion. Existence and uniqueness for such a
non-linear SDE is not at all obvious and still open.

\subsection{Local variance Gamma model}

A second example, introduced by P. Carr \cite{car}, which does not
require the construction of a continuous-time implied volatility
surface is given by the local variance Gamma model in which the
process $X_t$ is defined as a time-homogeneous one-dimensional It\^o
diffusion $\bar{X}_t$ subordinated by an independent Gamma process
$\Gamma_t$ \cite{car}:
 \b*
 X_t \equiv \bar{X}_{\Gamma_t} \\
 d\bar{X}_t=\sigma(\bar{X}_t) dW_t,\; \bar{X}_0=X_0
 \e*
The distribution of the Gamma process at time $t$ is a Gamma distribution with density:
 \b*
 \P\{\Gamma_t \in ds\}
 &=&
 {\alpha^{t \over t^*} \over
 \Gamma\left({t\over t^*}\right)}
 s^{{t\over t^*}-1} e^{-\alpha s}
 \;,\; s>0
 \e*
for some parameters $t^*=t_2-t_1$, $\alpha=1/t^*$.
The Fokker-Planck PDE reads
 \b*
 {1 \over 2} \sigma(K)^2
 \partial_K^2 {\cal C}(t_2 ,K)
 &=&
 { {\cal C}(t_2,K) -{\cal C}(t_1,K)\over t_2-t_1}
 \e*
from which we can deduce the local volatility
function $\sigma(\cdot)$ from call options valued uniquely at $t_1$ and $t_2$. The Dupire infinitesimal
calendar spread gets replaced by a discrete calendar spread. Similar to the previous example, a rigorous
adaptation of this idea requires a regularization of the above function $\sigma(\cdot)$ as in \cite{hir0}.

\subsection{Local L\'evy's model} \label{LocalLevy}

As a last example, we review the local Levy model introduced by Carr, Geman Madan, and Yor \cite{cgmy}. The
process $X_t$ is a compensated jump martingale
 \b*
 dX_t
 &=&
 \int_\R X_{t-}\left( e^x-1\right)
          \left( m(dx,dt)-\nu(dx,dt)\right),
 ~~\nu(dx,dt)=a(t,X_t)k(x) dx dt,
 \e*
where $m(dx,dt)$ is the counting measure with compensator $\nu$.
The analogue of the Dupire formula is
 \b*
 \partial_t{\cal C}
 &=&
 \int_0^\infty \partial_y^2{\cal C}(t,y) y a(t,y)
               \psi\Big( \ln {K \over y} \Big) dy,
 \e*
with the double tail $\psi$ of the L\'evy measure $k(x)$ given by
 \b*
 \psi(t,z)
 &=&
 \1_{\{z<0\}}\int_{-\infty}^z e^{x} \int_{-\infty}^x k(u) du dx
 +\1_{\{z>0\}}\int_{z}^\infty e^{x}\int_{x}^\infty  k(u) du dx.
 \e*

\end{document}